\documentclass[jcp,article,accept,pdftex,moreauthors]{Definitions/mdpi} 

\firstpage{1} 
\pubvolume{6}
\issuenum{4}
\articlenumber{129}
\pubyear{2026}
\copyrightyear{2026}
\datereceived{16 June 2026} 
\daterevised{25 July 2026 } 
\dateaccepted{30 July 2026 } 
\datepublished{1 August 2026} 
\pdfoutput=1 

\Title{Assessing AI-Generated vs. Human-Authored Spear Phishing SMS Attacks: An Empirical Study}

\Author{Jerson Francia$^{1,*}$\orcidA{}, Derek Hansen$^{1,*}$\orcidB{}, Benjamin Schooley$^{1}$ \orcidC{}, Matthew Taylor$^{1}$, Shydra Valynn Murray$^{1}$ \orcidD{}, Rebekah~Cornelius$^{1}$ \orcidE{} and Greg Snow$^{2}$\orcidF{}}

\AuthorNames{Jerson Francia, Derek Hansen, Benjamin Schooley, Matthew Taylor, Shydra Valynn Murray, Rebekah Cornelius, and Greg Snow}

\address{%
$^{1}$ \quad Department of Electrical and Computer Engineering, Brigham Young University, Provo, UT 84602, USA;
 ben\_schooley@byu.edu (B.S.); mtaylo48@byu.edu (M.T.); shywilli@byu.edu (S.V.M.); corn07@byu.edu (R.C.)\\
$^{2}$ \quad Department of Statistics, Brigham Young University, Provo, UT 84602, USA; snow@stat.byu.edu (G.S.)}

\corres{\hangafter=1 \hangindent=1.05em \hspace{-0.82em} Correspondence: jersno@byu.edu (J.F.); dlhansen@byu.edu (D.H.)}

\abstract{Personalized phishing is difficult to defend against because messages can be tailored to a target's work, interests, and social context. Large language models may make such tailoring faster and easier, but it remains unclear whether messages produced from simple prompts are more convincing than those written by people. This 25-target pilot study compared personalized smishing messages generated by GPT-4 with messages written by novice student authors working under time constraints. Using the proposed Threshold Ranking Approach for Personalized Deception (TRAPD), participants ranked 12 messages written for them, indicated the point at which they would intend to click, explained their reasoning, and judged whether each message was authored by GPT-4 or a human. GPT-4-generated messages elicited an intention to click more often than student-authored messages (28\% versus 21\%), although the difference was uncertain. More broadly, our findings suggest that a simple prompt can produce personalized messages that participants found comparably convincing within the uncertainty of this pilot study. Job-related messages were significantly more likely to elicit an intention to click than hobby- or social-media-related messages. When asked whether a message was written by a human or generated by AI, participants identified the source no more accurately than chance, although the two study-specific message sets remained computationally distinguishable based on their text. Together, these findings suggest that accessible AI-assisted personalization may increase the practical scale of social-engineering threats, while also demonstrating both the value and current limitations of TRAPD for controlled and ethical comparison.}


\begin{document}

\section{Introduction}

\label{sec:introduction}
Cybersecurity is a dynamic battle between attackers and defenders continuously adapting to new and emerging technologies. Although~robust security measures can mitigate cyber risk, the~strength of a system often depends on its users~\cite{alsharida_systematic_2023}. Among~the myriad tactics exploited by threat actors, phishing remains the most common way to infiltrate systems. Phishing employs social engineering to persuade victims to perform an action, such as clicking on a malicious link or email attachment, that causes malicious code to run or disclose sensitive information. Phishing attacks usually take the form of email messages, telephone calls (vishing), or~SMS messages (smishing) \cite{yeboah-boateng_phishing_2014}. According to the FBI Internet Crime Complaint Center, phishing/spoofing remained the most frequently reported cybercrime category in 2025, with~191,561 complaints and \$215.8 million in reported losses~\cite{fbi_ic3_2025}.

\textls[-15]{Spear phishing is a targeted phishing method that uses contextual information about a victim or organization to create more credible deceptive messages~\cite{dewan_social_2014,mohamed_spear_2024,schmitt_digital_2024}. Targeted spear phishing poses a significant threat as its prevalence and sophistication continue to grow. The~Proofpoint ``State of the Phish 2024'' report shows that spear phishing remained prevalent in both 2022 and 2023, with~reported prevalence comparable to bulk phishing across those years~\cite{stateofphish_2024}. In~instances where the pretext, or~the attacker's story, aligns with what the victim expects, the~attack is more likely to succeed~\cite{benenson_unpacking_2017}. For~instance, users expecting a shipping update from a retailer are more likely to fall for phishing messages imitating such communications~\cite{rajivan_creative_2018}. The~effectiveness of spear phishing stems from personalization, and~recent research suggests that modern artificial intelligence (AI) models can automate parts of the workflow for creating targeted content at substantially lower marginal cost~\cite{schmitt_digital_2024,heiding_evaluating_2024}.}

Traditionally, crafting these targeted messages was less common because it required substantial time and effort to research a victim and produce a plausible pretext~\cite{rajivan_creative_2018}. However, recent work has shown that commercial Large Language Models (LLMs) can generate convincing phishing emails and websites even without performing adversarial techniques to ``jailbreak'' the model~\cite{roy_chatbots_2024}. The~ability of AI to generate messages that appear increasingly human-like (and with limited expertise) has heightened the need to understand how cybersecurity defenses should respond to these capabilities. While human-authored phishing may rely on individual creativity and context, AI-generated phishing can combine personalization with speed and~scale.

Despite growing concern about AI-enabled phishing, an~important gap remains in the literature. Prior studies have shown that AI can generate phishing content and that personalization can increase message effectiveness, but~there is still limited evidence comparing GPT-4-generated and human-authored spear phishing messages when both are tailored to specific individuals, especially in the SMS context. There is likewise limited evidence on whether targets can meaningfully distinguish AI-generated spear phishing messages from human-authored ones, what cues they rely on when making that judgment, and~whether the message text itself carries detectable signal about its~source.

This 25-target pilot study investigates the use of an LLM to craft personalized spear phishing SMS, or~\textit{{spear smishing}
}, messages. We compare messages generated by GPT-4 from a simple shared prompt with human-authored messages written from the same prompt by novice student authors with time constraints. We evaluate the messages using a proposed evaluation framework called the Threshold Ranking Approach for Personalized Deception (TRAPD). The~design combines within-target rankings and intended-click thresholds with qualitative explanations, human source judgments, and~a study-specific computational analysis of whether the GPT-4- and human-authored messages can be distinguished from their text. For~the remainder of this paper, ``human-authored'' refers to this screened, time-constrained novice student-author condition and does not represent professional social engineers or human authors~generally.

This comparison matters for more than academic curiosity. Defenders often rely on assumptions about what phishing messages look like, including assumptions that human-authored malicious content will contain different stylistic cues than AI-generated content. If~those assumptions no longer hold, then both user training and detection approaches may need to adapt. By~directly comparing AI- and human-authored spear smishing messages, this study helps clarify how current LLM capabilities affect the evolving threat~landscape.

This paper makes five contributions. First, it provides a controlled pilot comparison of GPT-4-generated and human-authored personalized smishing messages written for the same targets. Second, it measures whether targets can identify message source and documents the cues they used. Third, it identifies message features that participants associated with convincingness and suspicion, including relevance, sender identity, URLs, communication medium, style, and~contextual accuracy. Fourth, it tests whether the GPT-4- and human-authored messages in this study remain distinguishable from text after obvious surface cues are removed and message lengths are matched, using tests on previously unseen targets. Fifth, it introduces TRAPD as a proposed evaluation framework and documents its feasibility and limitations for controlled, within-target comparison of personalized deceptive content.

\section{Review of the Related~Literature}

The body of literature related to spear phishing is expanding, with~studies investigating factors that lead individuals to fall for phishing attacks~\cite{benenson_unpacking_2017,oliveira_empirical_2019,alsharida_systematic_2023,lin_susceptibility_2019} and the application of AI to detect and generate such attacks~\cite{hazell_large_2023,seymour_generative_2018}. Prior work shows that personalization matters, that phishing extends beyond email, and~that recent AI systems can generate phishing content at scale. Less is known about how current LLM-generated messages compare with human-authored messages when both are personalized to the same targets, particularly in SMS. Prior research has also not fully characterized how people judge whether personalized SMS messages were generated by AI or written by people. This section reviews the literature leading to these~gaps.

\subsection{Evolution of Phishing~Techniques}
As communication methods evolved, so did phishing techniques, expanding from email scams into targeted, multi-channel social-engineering attacks~\cite{althobaiti_review_2024,barrera_literature_2024,apwg_trends_2025}. Early phishing scams primarily relied on email because of its widespread use~\cite{alsharida_systematic_2023,karamagi_review_2021}. Poorly executed messages were often recognizable through spelling, grammar, or~header problems. Recent LLM-based phishing research suggests that these language cues are becoming less reliable~\cite{bethany_lateral_2025,roy_chatbots_2024}. Attackers can create {spear phishing} messages by incorporating contextual information about individuals or organizations~\cite{mohamed_spear_2024,schmitt_digital_2024}, including information from social media, public records, or~previous breaches~\cite{dewan_social_2014,mohamed_spear_2024,schmitt_digital_2024}.

While email remains a major channel, phishing now extends across SMS, social media, and~various messaging platforms~\cite{verizon_dbir_2025,mishra_sms_2019,bossetta_weaponization_2018}. SMS phishing, also known as {smishing}, takes advantage of our growing reliance on text messaging via mobile devices. Smishing exploits the ubiquity and immediacy of mobile messaging, often using urgency, fear, or~impersonation to deceive users~\cite{barrera_literature_2024,blancaflor_unmasking_2024}. Smishing messages often appear to come from trusted entities like delivery services, financial institutions, or~online platforms, since many legitimate companies use SMS to automate user alerts and~notifications.

Social media platforms have also made phishing easier by allowing attackers to impersonate others and gather information for more targeted scams~\cite{bossetta_weaponization_2018,althobaiti_review_2024,apwg_trends_2025}. Attackers can create fabricated profiles of real individuals or organizations to reach out to their targets or publish deceptive posts or ads~\cite{bossetta_weaponization_2018, parker_factors_2020}. The~open nature of SMS and social networks makes it difficult to combat phishing in these media. Despite the diversity of these attack vectors, comparatively little work has examined whether the source of a personalized message (AI-generated or human-authored) changes either its effectiveness or a target's ability to recognize it. This gap is especially relevant in short-form SMS contexts, where prior comparative studies have focused primarily on email-based or organization-wide phishing~\cite{bethany_lateral_2025,heiding_evaluating_2024}.

\subsection{AI-Enabled~Phishing}
The integration of machine learning (ML), neural networks, and~LLMs into cybercrime creates new opportunities for phishing by making it easier to automate tasks that require cognitive effort, creativity, and~persuasive language. In~a study by Zhai~et~al., models like GPT-4 were used to solve a variety of tasks, ranging from common-sense reasoning to more complex problem solving~\cite{zhai_problem_2023}. These findings suggest that LLMs can perform cognitively demanding and creative tasks, which helps explain their potential application to social~engineering.

Early AI-phishing studies helped establish the feasibility of automating persuasive malicious content, but~they differ in important ways from the present study. Seymour and Tully trained a neural network on social media posts to create spear phishing messages, demonstrating that AI could be used to create targeted attacks and reporting a 30-66\% success rate in their experimental context~\cite{seymour_generative_2018}. Khan~et~al. used GPT-2 to generate phishing attacks and combined that model with a game-theoretic approach to improve attack decision making~\cite{khan_offensive_2021}. These studies are important precursors, but~they do not provide a direct comparison between current AI-generated and human-authored spear phishing messages aimed at the same individual~targets.

Recent cybersecurity research frames generative AI as more than a persuasive text-generation tool. Schmitt and Flechais argue that generative AI strengthens social engineering through three mutually reinforcing capabilities: realistic content generation, advanced targeting and personalization, and~automated attack infrastructure~\cite{schmitt_digital_2024}. In~a related work, Roy~et~al. demonstrate that commercial LLMs can be used to generate convincing phishing emails and websites without requiring model modification or prior jailbreaking~\cite{roy_chatbots_2024}. Together, these studies suggest that the key concern is not only the quality of LLM-generated messages, but~also the reduced effort, expertise, and~cost required to produce spear phishing content at~scale.

Recent comparative studies have begun to evaluate whether LLM-generated phishing can match human-authored phishing in realistic settings. Bethany~et~al. examined LLM-generated lateral phishing in a large organizational context and found that LLM-generated emails were as effective as those written by communications professionals~\cite{bethany_lateral_2025}. Heiding~et~al. similarly evaluated LLM-enabled phishing in a human-subjects setting, comparing human-expert emails, fully automated AI emails, and~human-in-the-loop AI emails~\cite{heiding_evaluating_2024}. These findings concern email and do not settle whether the same pattern holds for personalized SMS messages written for the same~targets.

Computational source attribution also has important limits. Writers can deliberately revise AI-generated text to make it harder for detectors to identify. Cheng~et~al. found that detector-guided paraphrasing substantially reduced the effectiveness of several AI-text detectors~\cite{cheng_adversarial_2025}. Classifier results should therefore be read as findings about the messages and conditions tested, not as general indicators of AI authorship when writers can revise~outputs.

The comparison is important because human and AI authors may bring different strengths to spear phishing. Human-authored spear phishing traditionally relies on the author's ability to gather context, identify vulnerabilities, and~adapt social-engineering strategies to a target. However, manual spear phishing is difficult to scale because each message must be individually crafted. AI-generated phishing can reduce that scalability barrier by automating parts of reconnaissance and message generation~\cite{hazell_large_2023,seymour_generative_2018,heiding_evaluating_2024,schmitt_digital_2024}. Prior work also suggests that combining LLMs with other strategies can reduce some limitations in contextual nuance~\cite{khan_offensive_2021, seymour_generative_2018}, while the current generation of LLM-enabled phishing already requires new detection and prevention methods~\cite{wilczynski_resistance_2024}.

Taken together, the~literature shows that successful spear phishing depends heavily on personalization, that phishing has expanded beyond email into channels such as SMS, and~that AI systems are increasingly capable of generating convincing phishing content at scale. However, there remains limited direct evidence comparing current LLM-generated and human-authored messages when both are personalized to the same individual targets in the smishing context. There is also limited evidence about whether targets can distinguish between AI- and human-authored messages, what criteria they use when making that judgment, and~whether the message text itself carries detectable signal about AI-versus-human source. The~present study addresses this gap by comparing AI- and human-authored spear phishing SMS messages for the same targets and by analyzing both perceived convincingness and source~attribution.

\section{Research~Questions}

We address the following five research questions about personalized spear phishing SMS messages and human and computational responses to them:

\noindent\textbf{{RQ1.} 
} In this pilot study, how did participants rate the convincingness of GPT-4-generated and human-authored personalized smishing messages?

\noindent\textbf{{RQ2.}} What content characteristics contribute to a more convincing spear phishing message?

\noindent\textbf{{RQ3}.} How accurately did participants identify whether the study messages were GPT-4-generated or human-authored?

\noindent\textbf{{RQ4.}} What criteria did participants describe when judging whether the study messages were GPT-4-generated or human-authored?

\noindent\textbf{{RQ5.}} Can the GPT-4-generated and human-authored messages in this study be distinguished computationally from their text?

\section{Methodology}
This study uses a proposed evaluation framework we call TRAPD, which stands for Threshold Ranking Approach for Personalized Deception. TRAPD was developed for situations in which researchers need to compare personalized deceptive content across conditions while preserving informed consent and avoiding the ethical and procedural barriers associated with live deceptive deployment. In~this pilot, it is used to compare GPT-4-generated and human-authored spear phishing SMS messages tailored to the same targets, while also collecting qualitative feedback about participants' decisions. It does not reproduce a live phishing attack or directly measure real-world click~behavior.

\textls[-15]{The study contains two analytic tracks. The~first examines convincingness: which messages did participants find most and least convincing, and~why? The second examines source attribution: could participants and a computational analysis distinguish GPT-4-generated from human-authored messages in this study, and~what cues did participants~use?}

For this study, spear phishing SMS messages were created and evaluated by 25 targets. Participants provided informed consent, and~the study was reviewed and approved by the Brigham Young University IRB (protocol IRB2023-065).

\subsection{{TRAPD Framework and Study Procedure} 
}
One contribution of this paper is to introduce TRAPD as a proposed evaluation framework for comparing personalized deceptive messages in an ethical manner. Although~this study focuses on spear phishing SMS messages, the~framework could be adapted for other types of deceptive content tailored to individuals, such as personalized disinformation or other phishing channels. TRAPD supports comparison of different message conditions, such as messages created by humans and AI or messages using different topics. It combines quantitative and qualitative evidence so that the study examines not only which messages participants found more convincing, but~also how they explained those~differences.

TRAPD builds on controlled phishing studies, laboratory judgment tasks, and~personalized phishing simulations. Prior work has used surveys and laboratory tasks to study perceived convincingness, suspiciousness, and~intended responses to phishing messages~\cite{hoeken_perceived_2022,moody_phish_2017,zhuo_human-centered_2023,hakim_phishing_2021}, while other work has used controlled simulation environments to study how personal information can be exploited in spear phishing~\cite{xu_personalized_2023}. TRAPD adapts these approaches for cases in which researchers need to compare multiple personalized deceptive messages written for the same target while preserving informed consent. It therefore trades some ecological realism for ethical control, within-target comparison, and~richer explanations of participant~reasoning.

TRAPD is proposed as an evaluation framework for comparing multiple personalized deceptive messages written for the same target. In~this study, it supports internally valid within-target comparisons under controlled conditions, but~it has not yet been formally evaluated for reliability, agreement across evaluators, order effects, or~validity against live behavior. This study therefore shows how the framework can be used and identifies lessons for future studies rather than validating it as a general measurement~instrument.

At its core, the~TRAPD evaluation framework includes the following~steps:
\begin{enumerate}[leftmargin=*,labelsep=5mm]
    \item Recruit targets who willingly share personal information with potential attackers.
    \item Generate personalized deceptive messages aimed at the targets, for~example using humans or AI.
    \item Have targets rank the messages from most to least convincing and choose a threshold above which they would intend to click.
    \item Have targets explain why they placed messages where they did.
    \item Optionally, have targets label messages with a variable of interest, such as perceived AI authorship, and~explain their choices.
\end{enumerate}

\begin{figure}[H]
     \includegraphics[width=0.85\textwidth,keepaspectratio]{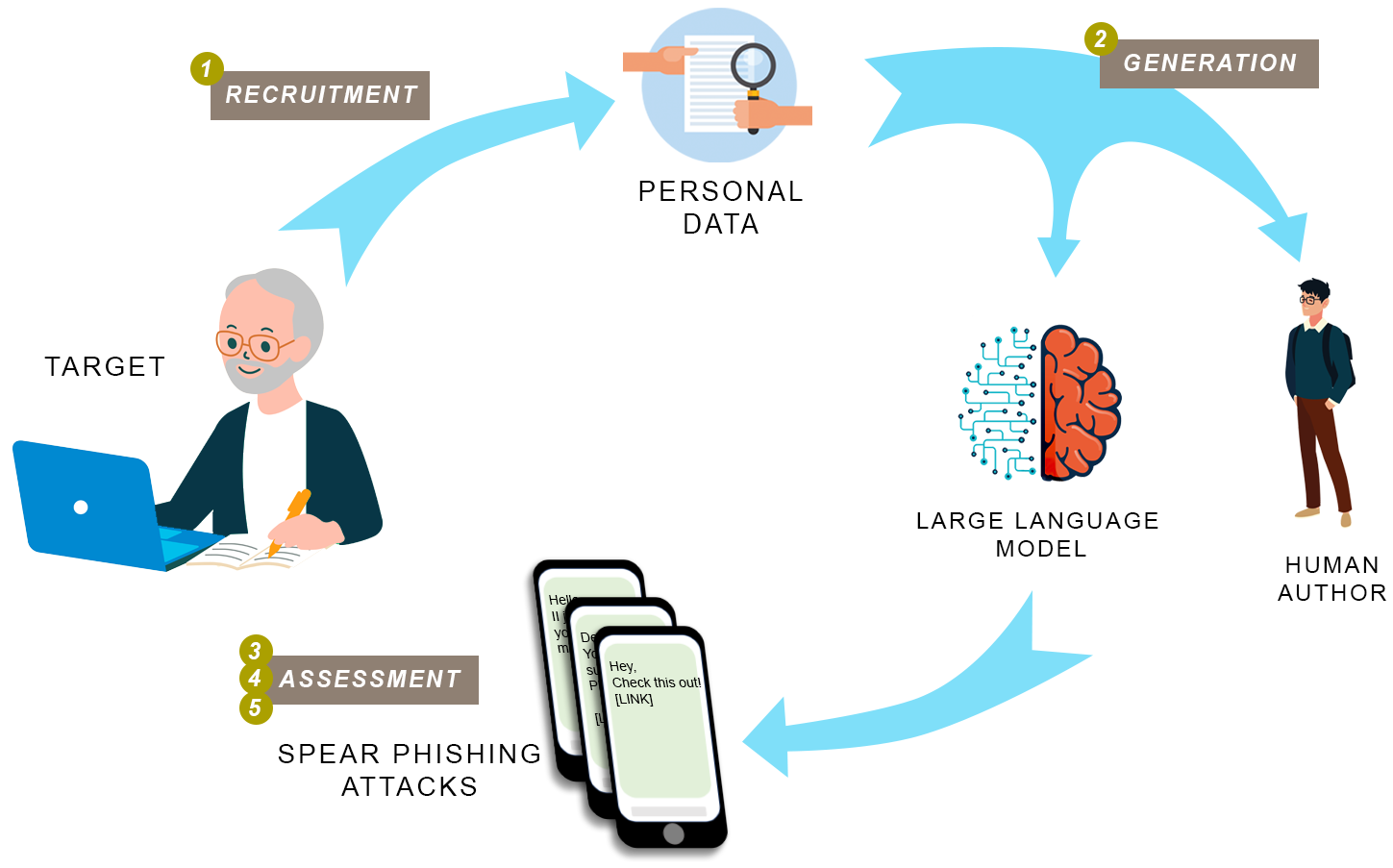}
    \caption{\label{fig_overview}{Overview of the study using the proposed TRAPD evaluation framework. Recruitment, generation, and~assessment are numbered according to the framework steps. Blue arrows indicate the study flow; numbered gold circles mark the TRAPD steps.} 
}
\end{figure}

Figure~\ref{fig_overview} illustrates the study flow according to these steps. In~this pilot, TRAPD served three practical purposes. It let us compare GPT-4-generated and human-authored messages written for the same person, link participants' rankings to the point at which they said they would click, and~collect their explanations for how they ranked and labeled the messages. These features support controlled within-target comparisons and help explain participant reasoning, but~they do not estimate real-world click-through~rates.

\unskip

\subsection{Recruiting Targets Who Shared Personal~Information}
The first phase recruited participants as potential targets and collected personal information through a survey. Recruitment used convenience channels, including people known to the researchers, printed flyers in the university library, posts on personal social-media accounts, and~email lists for university departments and neighborhoods. Interested individuals completed a Qualtrics survey, provided consent, and~shared information for generating personalized messages. They were told that they would later be invited to evaluate messages created for them, but~not whether the messages would be written by humans or~AI.

Forty-one people completed the survey. It collected their name, email address for interview scheduling, sex, age group, technical-ability rating, hobbies, city and state, occupation and workplace, and~an item from their home that they had recently posted on social media. Recruitment was not intended to produce a sample representative of the broader population. Of~these 41 people, 25 returned for the later interview and make up the final target sample. The~remaining 16 provided survey information but did not complete the ranking, intended-click, source-attribution, or~interview activities. The~interview completion rate was therefore 61\% (25/41), and~the findings are limited to the 25 participants who completed the~pilot.

\subsection{Generating Personalized Deceptive~Messages}
Using the survey information, GPT-4 and human authors generated spear phishing messages in SMS form. Each prompt used one of three topics: the target's hobbies, workplace, or~something the target owned and had recently posted about on social media. These topics represent common personalization sources rather than all possible phishing themes. Job prompts represent workplace pretexts~\cite{hanus_phish_2022,burns_phishing_2019,williams_exploring_2018}; hobby prompts represent personal interests~\cite{lin_susceptibility_2019,benenson_unpacking_2017,rajivan_creative_2018,xu_personalized_2023}; and social-media prompts represent publicly or semi-publicly shared information that can increase familiarity and perceived relevance~\cite{bossetta_weaponization_2018,parker_factors_2020,seymour_generative_2018}. Using the same three topics for each target kept the type of personalization consistent across the two message~sources.

\begin{figure}[H]
\hspace{-12pt}    \includegraphics[width=\linewidth,keepaspectratio]{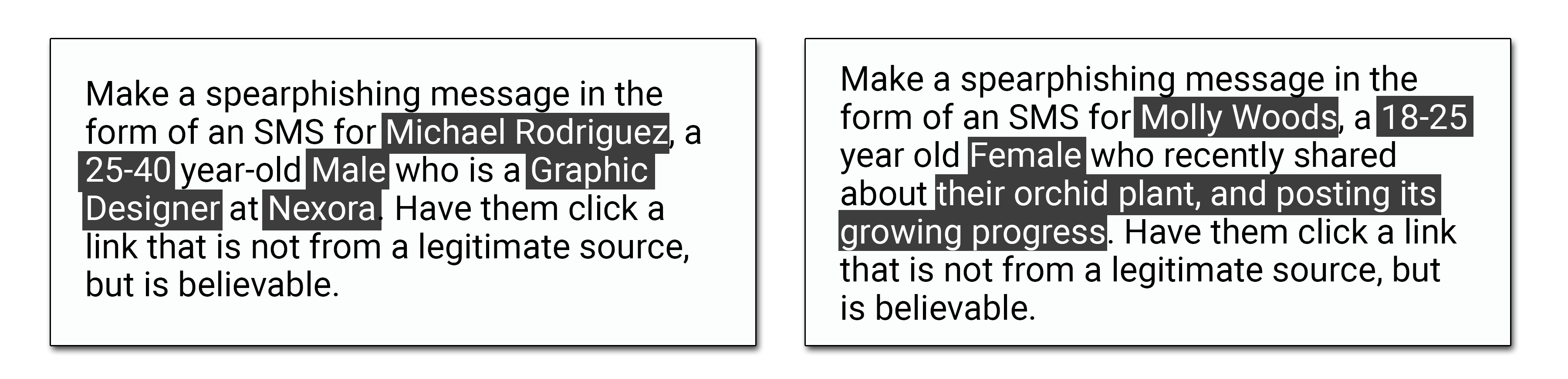}
    \caption{\label{fig_sample_prompt}{Pseudonymized examples of the prompt template.} \textls[-15]{The displayed names and organization are pseudonyms. Black-highlighted text marks target-specific fields inserted from the participant~survey.}}
\end{figure}

Figure~\ref{fig_sample_prompt} shows pseudonymized examples of the shared prompt structure. The~displayed personal names and organization are pseudonyms. Black highlighting identifies fields populated with target~information.

\vspace{-6pt}

\textls[-15]{The same prompt structure was provided to human authors and GPT-4. This supported a controlled comparison while allowing both conditions to compose messages~naturally.}

\subsubsection{Human~Generation}
The human-authored messages were created by time-constrained novice student authors rather than professional social engineers. The~99 authors were undergraduate cybersecurity students or honors students enrolled in a deception course. They had received phishing-related instruction and practice and were asked to write up to four messages in approximately 15 min. Students received extra credit for~participating.

Authors entered messages in an online survey tool that supported emojis and used a text box approximating an SMS message. They received up to four randomly selected prompts and were asked to use the supplied information to induce the target to click a link. After~each prompt, authors rated their confidence that the message ``would trick the target'' on a five-point scale and answered demographic and experience~questions.

\textls[-15]{The student authors submitted 363 messages, of~which 246 (67.8\%) passed the screening criteria. A~research team that included two cybersecurity professors reviewed the messages to establish a minimum quality threshold. Messages were excluded if they were incomplete, lacked a request to click a link, or~were clearly unusable. Generic shortened links replaced link placeholders. At~least three candidate messages were collected for every target/topic combination, and~two valid messages were randomly selected when more than two were available. This review improved the quality of the human-authored pool, but~it did not make the messages representative of those written by experts or real-world~attackers.}

\subsubsection{GPT-4~Generation}
The study's batch-generation notebook submitted the same target/topic prompt structure to the OpenAI API and collected three responses per prompt using the \texttt{gpt-4} model alias, temperature 0.7, and~a maximum output length of 256 tokens. Two outputs per topic were selected at random. After~both generation phases, each target had 12 messages: six GPT-4-generated and six human-authored, with~two from each source for each of the three~topics.

The study documentation does not record the exact GPT-4 model snapshot, API logs, complete package environment, a~single end-to-end runner, or~parameters and defaults not explicitly recorded in the notebook. The~documented procedure and evaluated messages are available, but~the original generation run cannot be recreated exactly. We did not generate a new message set with a current model because doing so would test a different model version rather than reproduce the condition used in this~study.

\textls[-15]{Future studies should record the provider, exact model and snapshot, access date, complete prompts, temperature, top-p, maximum tokens, number of samples, seed when supported, API version, retry behavior, content-filter events, and~all output-curation~decisions.}

\subsection{Target Interview and Sorting~Activity}
The 25 interview participants were split approximately evenly by gender: 48\% male and 52\% female. Thirty-six percent were between 18 and 25 years old, 40\% were between 26 and 40, and~24\% were over 40. Fifteen participants (60\%) were affiliated with the university, including eight students (32\% of the sample). Occupations included librarian, software engineer, instructional designer, sales agent, and~teaching~assistant.

\subsubsection{Threshold Rank~Order}
\begin{figure}[H]
    \centering
    \includegraphics[width=\linewidth,keepaspectratio]{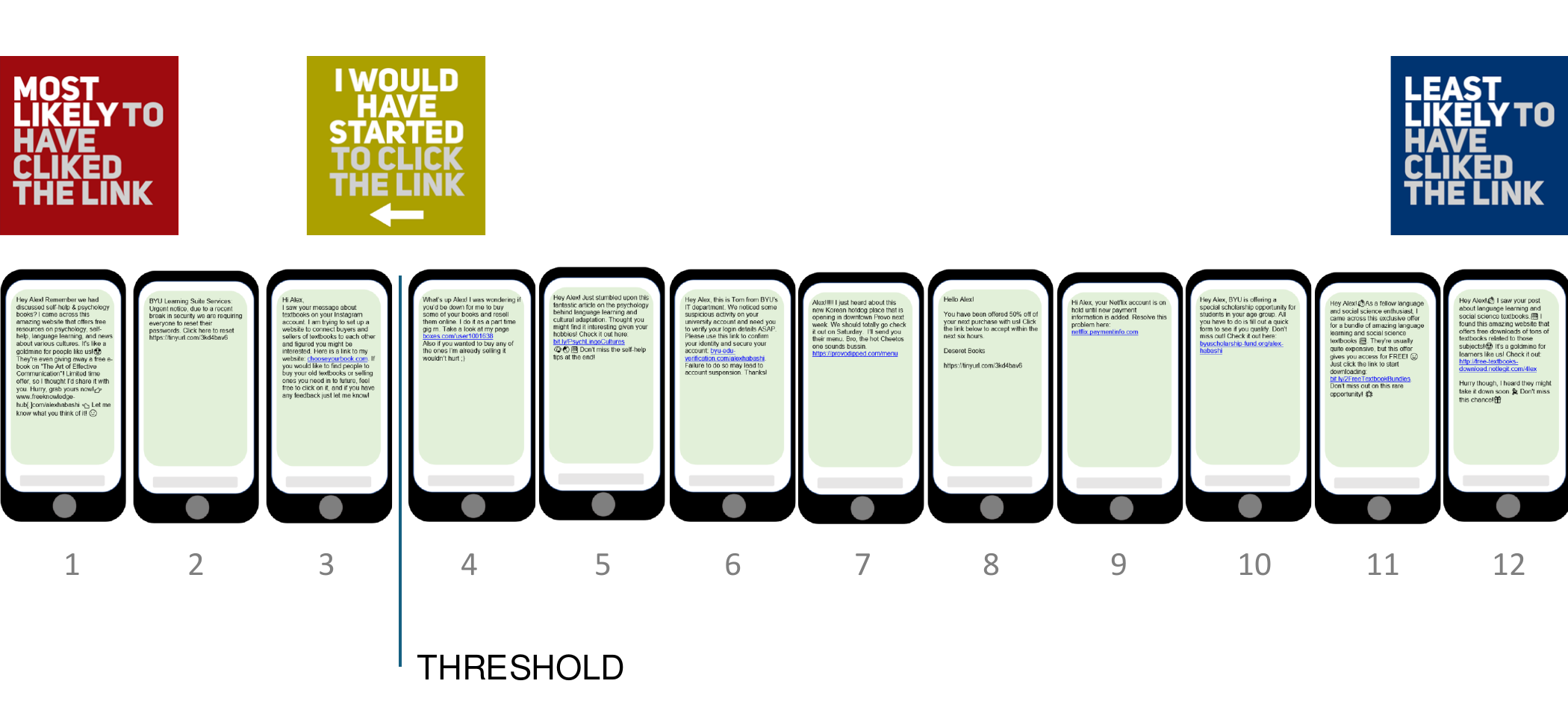}
    \caption{\label{fig_interview1}{Example of the threshold-ranking activity. Messages are ordered from most to least convincing, and~the marker indicates the participant's intended-click threshold. The text within the cards is illustrative; the arrangement, ranks, and threshold marker are the intended visual focus.} 
}
\end{figure}

Each participant received 12 printed representations of SMS messages. Participants arranged them from most to least likely to lead them to click and marked the point above which they would intend to click. They were not initially told that the messages were simulated spear phishing messages. Figure~\ref{fig_interview1} illustrates the~activity.

\unskip

\subsubsection{Qualitative~Assessment}
Participants explained what made messages seem convincing or suspicious and why they placed particular messages near the top or bottom. They were encouraged to ground their explanations in individual messages before discussing broader criteria~\cite{kahlke_when_2025}.

\subsubsection{Source~Labeling}
\begin{figure}[H]
    \centering
    \includegraphics[width=\linewidth,keepaspectratio]{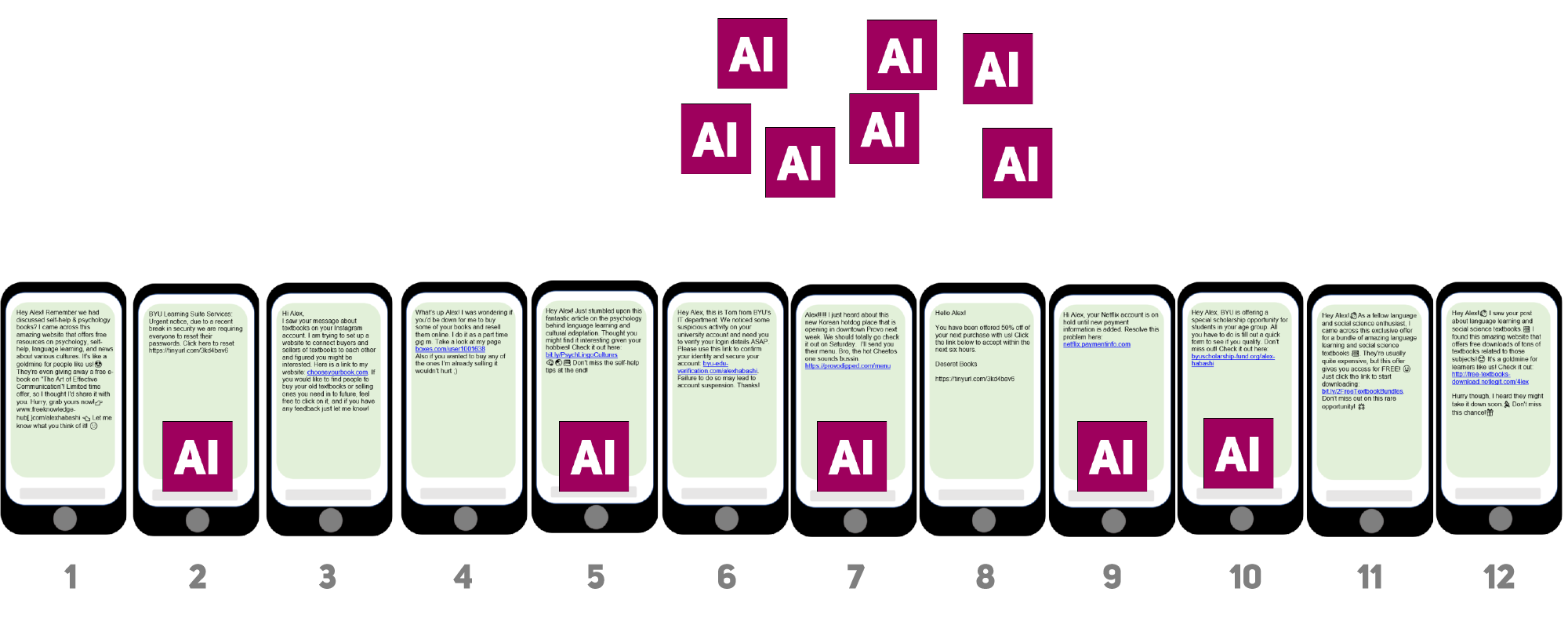}
    \caption{\label{fig_interview2}{Example of the source-labeling phase. Participants marked messages they believed were AI-generated after completing the earlier convincingness activities. Magenta `AI' labels show the participant's source judgments.} 
}
\end{figure}

After completing the ranking, intended-click threshold, and~convincingness explanation, participants were told that ``one or more of the messages were created by an AI.'' They then marked messages they believed were AI-generated and explained their human- or AI-authorship judgments. At~the participant's request, researchers revealed the actual sources afterward. All 25 interviews were audio recorded, and~the sorting and labeling arrangements were~photographed. Figure~\ref{fig_interview2} illustrates this source-labeling activity.

\vspace{-6pt}

Participants completed the phases in a fixed order. Learning about AI could not change the earlier rankings or intended-click thresholds, but~it may have influenced the later source labels and explanations. The~in-person interviewer and participants' awareness that they were being studied may also have encouraged answers they thought the researchers~expected.

\subsection{Convincingness Analytic~Track}
The first analytic track addressed RQ1 and RQ2. To~address RQ1, we used two complementary measures of convincingness. First, we examined whether participants rated a message above their own intended-click threshold. We compared the estimated intended-click probability for GPT-4-generated and human-authored messages while accounting for the fact that each participant evaluated multiple messages. We report the estimated probability for each source, their difference, and~95\% confidence intervals. The~primary model used a logistic generalized estimating equation (GEE), which accounts for repeated judgments within each target. We also conducted an exploratory model that considered message topic and actual text length. This reporting hierarchy was refined during revision and was not~preregistered.

Second, we examined participants' rankings. For~each target, we compared the average rank of the GPT-4-generated messages with that of the human-authored messages; a lower numerical rank indicates a higher ranking and greater perceived convincingness. We used a target-level sign-flip test for the within-target difference. Bootstrap confidence intervals, paired $t$-tests, and~Wilcoxon signed-rank tests provide additional checks on the intended-click comparison rather than additional primary~tests.

To address RQ2, we used the same measures to examine whether topic mattered. A~participant-clustered GEE compared intended clicking for job, hobby, and~social-media-related messages, followed by three pairwise contrasts. We also compared topic-specific average ranks within targets. Holm adjustment accounted for the three comparisons in each family. Demographic and prior-experience analyses were exploratory because the 25-target sample was too small for stable subgroup~conclusions.

To address RQ2 qualitatively, we examined interview transcripts to understand why messages felt convincing or suspicious. Recordings were automatically transcribed with Otter.ai. Primary coders checked their assigned transcripts against the recordings and corrected transcription errors. Names were replaced with pseudonyms or target identifiers, access to working transcripts was limited to the research team, and~quotations selected for publication were screened for direct~identifiers.

We used primarily inductive thematic coding, meaning themes were developed from participants' comments rather than imposed in advance. Three research-team members served as primary coders for the convincingness and source-attribution analyses, with~two additional team members providing supervision. Each transcript had one primary coder and was not independently coded in full by a second primary coder. For~RQ2, coders identified meaningful passages and message-specific comments explaining why a message seemed more or less convincing. During~team review, researchers compared proposed themes, combined overlapping categories, clarified definitions, revisited earlier transcripts when definitions changed, and~resolved differences through discussion and consensus. A~passage could receive more than one code. We did not calculate a formal inter-coder agreement~statistic.

\subsection{Source Attribution Analytic~Track}
The second analytic track addressed RQ3--RQ5. To~address RQ3, we measured how often participants correctly identified whether each message was GPT-4-generated or human-authored. Each participant made 12 judgments, for~a total of 300. Because~each participant judged multiple messages, the~primary accuracy analysis used a participant-clustered intercept-only GEE. We also resampled targets and compared estimated accuracy with the 50\% expected from random~guessing.

To explore RQ4 quantitatively, we examined whether correct source judgments were associated with emojis, modified links, or~message length. This analysis can identify patterns in these judgments but cannot establish that a feature caused a source judgment. Qualitatively, we used a separate set of inductive themes because source and convincingness judgments answer different questions, but~followed the same coding process. Theme percentages indicate how many of the 25 targets mentioned a theme at least once. Interviewers used open-ended questions and did not request comments on specific features. Quotations are verbatim except for anonymization, transcription corrections, and~bracketed~clarification.

\subsubsection{{Computational Source-Attribution Analysis Plan} 
}
To address RQ5, we asked whether the GPT-4-generated and human-authored messages in this study could be distinguished from their text alone. The~analysis used the same 300 messages shown during the interviews: 150 per source, balanced across targets and topics. It tests whether these study-specific message sets were distinguishable, not whether arbitrary AI-generated text can generally be~detected.

We converted each message into a semantic embedding using the OpenAI {text-embedding-3-small model} 
 and trained an L2-regularized logistic regression classifier. We used target-grouped five-fold cross-validation. In~each fold, the~model was trained on messages from 20 targets (80\%) and tested on messages from five unseen targets (20\%); messages from the same target never appeared in training and testing. We report pooled out-of-fold ROC AUC, average precision, balanced accuracy, sensitivity to GPT-4-generated messages, and~specificity to human-authored~messages.

Because the two message sets differed in visible features such as URLs, emojis, and~length, we conducted sensitivity analyses to test whether the classifier could still distinguish them after those features were controlled. We first standardized URLs, removed emojis, replaced digits with a common token, converted text to lowercase, removed punctuation and symbols, and~normalized whitespace. The~strictest condition also paired GPT-4-generated and human-authored messages within target/topic combinations and shortened both to the length of the shorter message. This retained all 300 messages while equalizing the source-specific length distributions. Participants saw only the\linebreak original~messages.

To protect participant privacy, the~additional embedding analyses used only normalized text; the original personalized messages were not sent to the embedding service. Supplementary Materials, {Section~1,} 
 reports an earlier analysis of messages with links removed. Because~that analysis used embeddings created at a different time, its results provide context but cannot be compared directly with the present sensitivity analyses to isolate the effect of any one~feature.

\subsubsection{PCA and Exploratory~Clustering}
To give a visual overview of which messages had similar or different text patterns, we projected the no-link message embeddings onto their first two principal components. We also explored whether the messages formed groups with similar text patterns by varying the number of principal components, clusters, and~random starting points. The~resulting exploratory KMeans analysis is reported in Supplementary Materials, {Section~2}.

\subsection{Implementation and Reproducibility~Details}
The original statistical analyses were conducted in R. The~final statistical and computational analyses were run in Python 3.13.12 using NumPy 2.5.1, pandas 3.0.3, SciPy 1.18.0, statsmodels 0.14.6, and~scikit-learn 1.9.0. Analyses involving randomness used seed 42. The~available participant-level data were sufficient to reproduce the reported results and requested sensitivity~analyses.

The available generation outputs account for all 150 GPT-4-generated messages evaluated in the study. However, the~exact model snapshot, API logs, complete software environment, a~single end-to-end generation runner, and~unspecified parameter defaults were not~recorded.

A repository accompanying this manuscript will provide final analysis code, aggregate outputs, fictionalized or placeholder-only prompt documentation, package versions, and~execution instructions. Personalized messages, participant-level data, historical notebooks, transcripts, recordings, and~embeddings will not be released because they contain privacy-sensitive material or could create dual-use~risks.

\section{Convincingness~Results}
This section addresses RQ1 and RQ2 by presenting the quantitative and qualitative results related to message~convincingness.

\subsection{Quantitative Convincingness~Results}
The main descriptive measures for each topic and source are average rank and intended click rate. Messages ranked closest to the top received the lowest numerical rank (1~=~highest rank and most convincing; 12 = lowest rank and least convincing), and~intended click rate was determined by the threshold marker. Table~\ref{table_rank_click} summarizes both~measures.

\begin{table}[H]
\centering
\caption{{Average rank and intended click rate by topic and source.} A lower average rank number indicates a higher ranking and greater perceived convincingness. Values are descriptive; the source difference in intended clicking was not statistically reliable in the participant-clustered analysis ($p = 0.147$). Topic-specific inferential results are reported separately in the~text. Bold labels separate the topic and source blocks.}
\begin{tabularx}{\textwidth}{lCC}
\toprule
\textbf{Category} & \textbf{Avg. Rank} & \textbf{Intended Click Rate} \\
\midrule
\multicolumn{3}{l}{\textbf{{Topic}}} \\ 
{Job} 
    & {5.71} & {38\%} \\
Hobby  & 6.66 & 19\% \\
Social & 7.13 & 17\% \\
\midrule
\multicolumn{3}{l}{\textbf{{Source}}} \\ 
Human-authored & 6.59 & 21.3\% \\
GPT-4-generated & 6.41 & 28.0\% \\
\bottomrule
\end{tabularx}
\label{table_rank_click}
\end{table}
\unskip

\subsubsection{GPT-4 vs. Human Ranking and Intended Click~Probability}
To assess the convincingness of GPT-4-generated and human-authored messages (RQ1), we examined both intended clicking and participants' message~rankings.

\begin{figure}[H]
\hspace{-6pt}    \includegraphics[width=0.9\columnwidth]{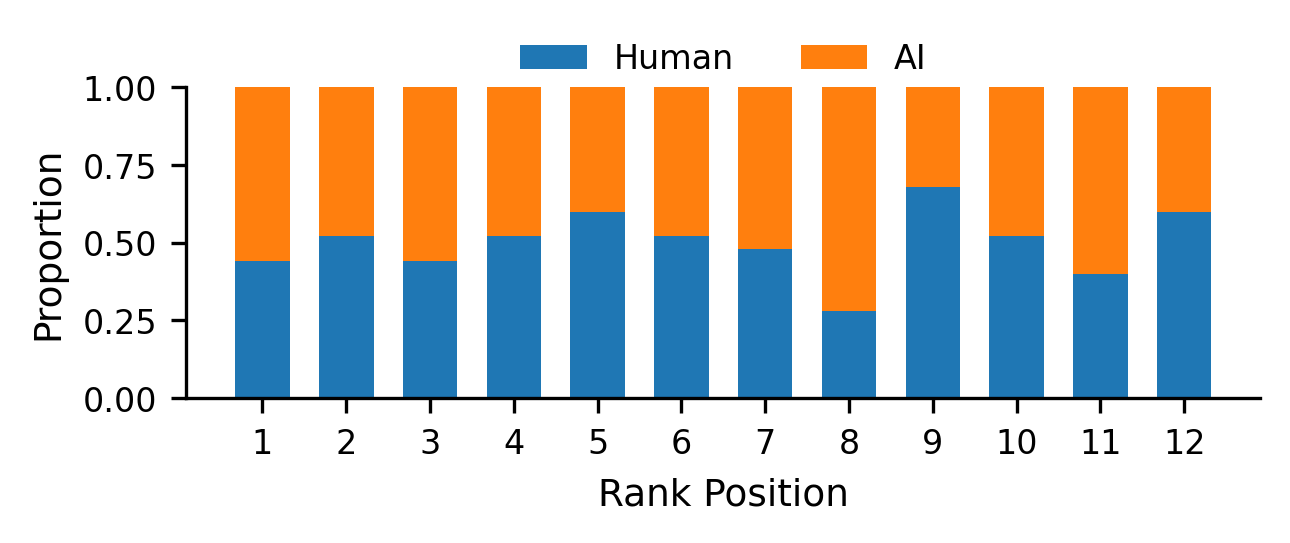}
    \caption{{Rank distribution by message source. GPT-4-generated and human-authored messages had similar rank distributions; the within-target rank difference was not statistically reliable.} 
}
    \label{fig:rank_distribution}
\end{figure}

Figure~\ref{fig:rank_distribution} shows the distribution of ranks by message source.

\vspace{-9pt}

Participants placed 28.0\% of the GPT-4-generated messages and 21.3\% of the human-authored messages above their intended-click thresholds. The~estimated difference was 6.7 percentage points. However, the~95\% confidence interval ranged from 2.9 percentage points in favor of human-authored messages to 16.3 points in favor of GPT-4-generated messages. The~model, which accounted for the multiple messages evaluated by each participant, therefore did not establish a reliable source difference (OR = 1.43, 95\% CI: 0.88--2.33; $p = 0.147$).

The rankings showed a similarly small difference. GPT-4-generated messages had a mean rank of 6.41, compared with 6.59 for human-authored messages; the smaller rank number means they were ranked slightly higher. This average difference of 0.17 rank positions was not statistically reliable in the within-target comparison ($p = 0.754$).

We also conducted an exploratory analysis that accounted for message topic and actual text length. After~considering these characteristics, the~estimated source difference was smaller and remained uncertain (OR = 1.16, 95\% CI: 0.71--1.91; $p = 0.551$). Longer messages were associated with a higher probability of intended clicking (OR = 1.30 per 100 additional characters; $p = 0.019$). Because~message length was not experimentally assigned, this association does not show that making a message longer would cause someone to intend to~click.

\begin{figure}[H]
    \includegraphics[width=\linewidth,keepaspectratio]{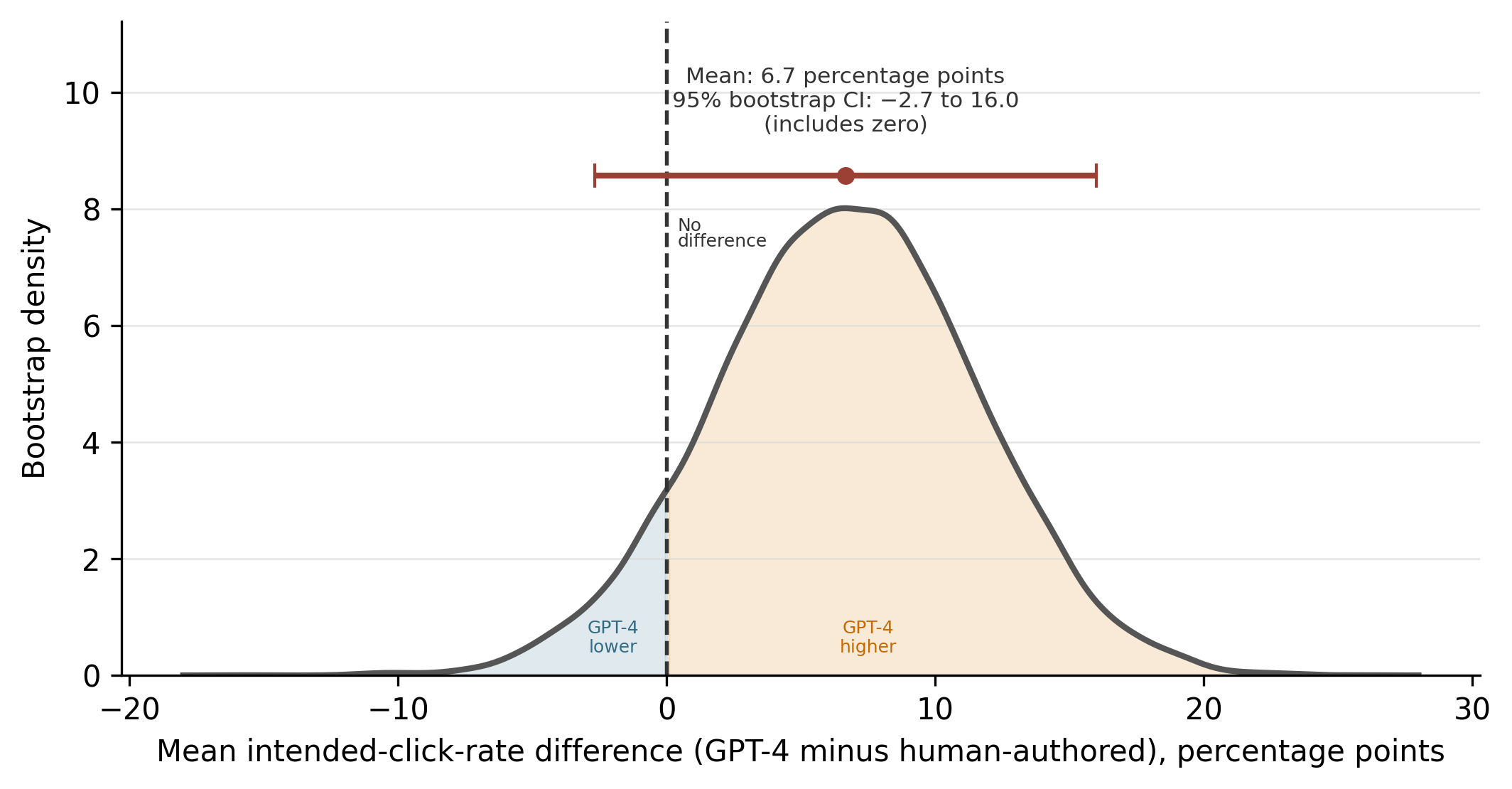}
    \caption{\label{fig_bootstrap}{Bootstrap distribution of the target-level mean intended-click difference (GPT-4 $-$ human-authored). The dashed vertical line marks no difference between sources. The mean difference was 6.7 percentage points, and~the 95\% confidence interval ranged from $-2.7$ to 16.0 percentage points. Because~the interval includes zero, the~study did not establish a source difference.} 
}
\end{figure}

Finally, we compared the two message sources separately within each target. Twelve targets had a higher intended-click percentage for GPT-4-generated messages, seven had a higher percentage for human-authored messages, and~six were tied. The~average within-target difference was again 6.7 percentage points, but~its bootstrap 95\% confidence interval ranged from $-2.7$ to 16.0 percentage points. Additional paired tests likewise did not find a statistically reliable source difference (paired $t$-test, $p = 0.195$; Wilcoxon signed-rank test, $p = 0.092$; sign-flip test, $p = 0.156$). Overall, the~results do not establish that either source produced more convincing messages, nor do they establish that the two sources were~equivalent. Figure~\ref{fig_bootstrap} shows the bootstrap distribution of the target-level intended-click difference.

\unskip

\subsubsection{Topic-Based Ranking and Intended Click~Probability}
\begin{figure}[H]
\hspace{-6pt}    \includegraphics[width=0.9\columnwidth]{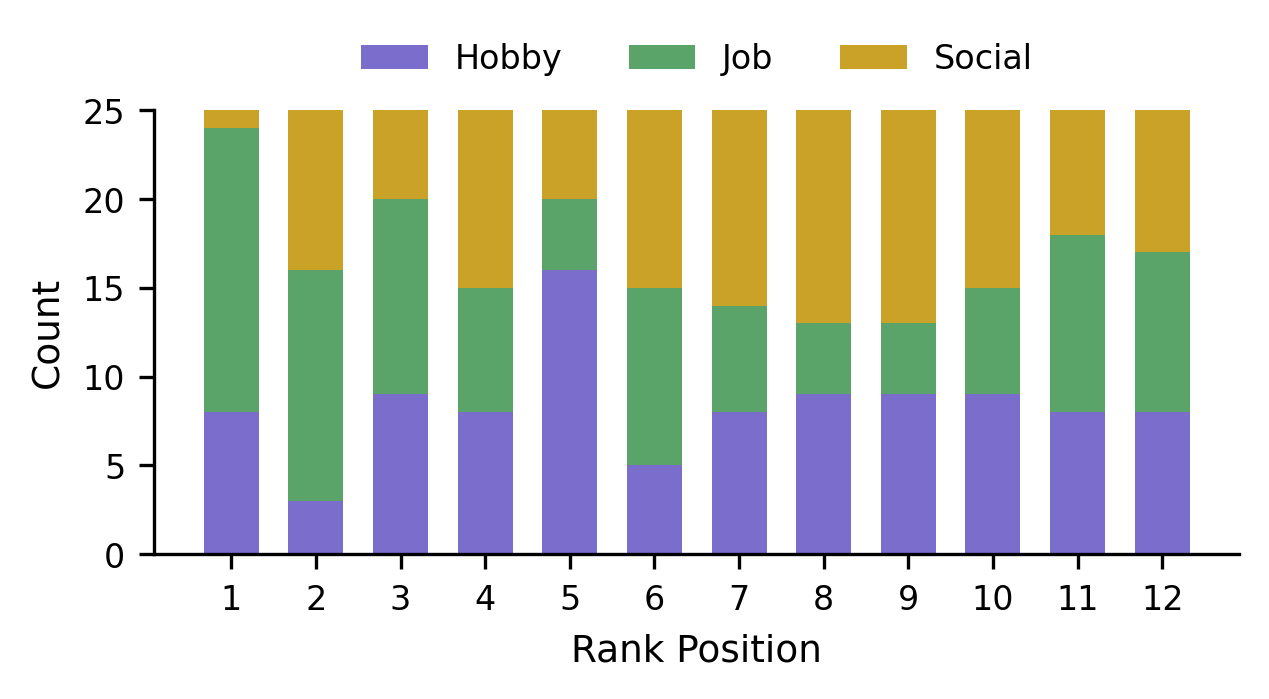}
    \caption{\label{fig_topicrank}{Rank distribution by message topic. Job-related messages were ranked highest on average descriptively, but~no pairwise rank comparison remained statistically reliable after Holm adjustment.} 
}
\end{figure}

RQ2 examined whether participants' responses differed by message topic: job, hobby, or~social media. Descriptively, job-related messages were the most convincing on both measures. Participants placed 38\% of job-related messages above their intended-click thresholds, compared with 19\% of hobby messages and 17\% of social-media-related messages. Job-related messages were ranked highest on average (mean rank = 5.71), followed by hobby messages (6.66) and social-media-related messages (7.13). Figure~\ref{fig_topicrank} shows the rank distributions by topic.

\vspace{-6pt}

The intended-click analysis, which accounted for the multiple messages evaluated by each participant, found an overall relationship between topic and intended clicking (Wald $\chi^2(2) = 10.85$, $p = 0.0044$). After~Holm adjustment for three comparisons, intended-click odds were higher for job-related messages than for hobby messages (OR = 2.61, 95\% CI: 1.16--5.89; adjusted $p = 0.0412$) and social-media-related messages (OR = 2.99, 95\% CI: 1.50--5.97; adjusted $p = 0.0056$). Hobby and social-media-related messages did not differ reliably (OR = 1.15, 95\% CI: 0.51--2.57; adjusted $p = 0.7427$).

Although the average rankings followed the same descriptive order, none of the pairwise rank differences remained statistically reliable after adjustment: job versus hobby, $p = 0.493$; job versus social media, $p = 0.147$; and hobby versus social media, $p = 0.493$. Thus, the~intended-click results support a difference by topic, particularly for job-related messages, while the ranking differences should be interpreted~descriptively.

\subsubsection{Demographics and Phishing~Experience}
The 25-target pilot sample was not large enough for stable demographic or prior-experience subgroup inference. These comparisons were treated as exploratory and are not interpreted as evidence that such differences are~absent.

\subsection{Content Characteristics of Convincing~Messages}
To address RQ2 qualitatively, we examined how participants explained why messages seemed more or less convincing, regardless of whether the messages were GPT-4-generated or human-authored. We grouped their explanations into recurring themes describing features they associated with convincingness or suspicion. These themes reflect participants' interpretations and should not be read as experimentally tested effects. Table~\ref{table_theme_convincing} reports the percentage of the 25 participants who mentioned each theme at least once. Participants could mention more than one theme and could describe the same feature as convincing in one message but suspicious in~another.

\begin{table}[H]
\centering
\caption{{Message features participants associated with convincingness or suspicion.} Percentages indicate the proportion of the 25 participants who mentioned each theme at least once and are not mutually~exclusive.}
\label{table_theme_convincing}
\begin{tabularx}{\textwidth}{lR}
\toprule
\textbf{Theme} & \textbf{\% of Participants} \\
\noalign{\hrule height 0.5pt}
{Relevance}    & {76\%} \\
Sender       & 68\% \\
URL          & 64\% \\
Medium       & 40\% \\
Style        & 40\% \\
Urgency and Scarcity & 32\% \\
Inaccuracies & 28\% \\
Rewards      & 28\% \\
\bottomrule
\end{tabularx}
\end{table}
\unskip

\subsubsection{Personal~Relevance}
Personal relevance was discussed by 19 of the 25 participants (76\%). Seventeen described relevant details as making some messages more convincing, while 17 also described irrelevant, inaccurate, or~insufficiently specific details as making other messages suspicious. Several participants pointed to messages directly tied to their occupational responsibilities, with~one saying that ``because that is my job is to help people with records...this is one that I feel like I'm most likely to engage with'' (T13). Another target, who is a banker at a credit union, noted the similarities between the targeted message and messages they receive at work. They noted that the ``alert literally looks like the alert we get [at work] when there's a fraud'' (T33). Some participants pointed out how the spear phishing messages aligned with their personal interests. One expressed enthusiasm about a message describing a ``Vineyard gardening club,'' believing that ``somebody from our community'' may have sent it. They expressed that they ``would love to get involved in that'' (T31). One more skeptical participant mentioned that although ``this interests me and it could exist, but~I'm not gonna click on it'' (T08).

On the other hand, personal relevance was perceived as unconvincing to 17 of the targets. Some participants highlighted messages that didn't align with their current activities. For~example, one interviewee noted, ``I'm also not actively dancing anymore, so that's just weird that they're offering me something like this'' (T25). Another target was suspicious of a social media-related message because they ``don't have an Instagram [account]'' (T34). Some targets emphasized messages that failed to pique their interest, with~one saying that they are ``not into Brandon Sanderson [the author]'' (T37). Another mentioned that they have already rescued an animal and ``don't need more right now'' (T02). Some targets mentioned messages that contained information that they had not disclosed. One target noted that the message wasn't ``legit'' because they ``never put [their] studio equipment online'' (T31), while another mentioned that ``it's a little weird to be selected for something that you don't apply for'' (T16).
Finally, targets showed skepticism when messages were somewhat relevant but lacked specificity. One target noted that ``there just wasn't enough content in the message body that was specifically directed at me'' (T20).

Because all messages were personalized, relevance was naturally a prominent theme. Participants generally associated accurate and specific personalization with greater credibility, while mismatched or vague personalization raised~suspicion.

\subsubsection{Sender~Identity}
Seven participants (28\%) described a familiar or plausible sender as making a message more convincing. One said the sender ``introduced herself as [their] neighbor'' (T30). Another noted, ``The sender started with 'My name is John''' (T41). One participant recognized ``actual company branding'' that resembled promotional texts they had received (T18), and~another said, ``This could be someone from my ward [church congregation], wanting me to like check out some plants or something'' (T31).

In contrast, 17 participants (68\%) described an unfamiliar or inconsistent sender as a reason for suspicion. One explained, ``James is my boss... it'd be weird that he was introducing himself that way... that doesn't sound like my boss'' (T31). Others asked how a sender obtained their number: ``So like, how did you get my number in the first place?'' (T27) and ``Like why would Lowe's have my phone number?'' (T20). T34 explained that an unrecognized number would prompt them to ask who was writing, while T31 said a message without a personal connection ``just seems like spam.'' Because the study messages did not display source phone numbers, these comments primarily concern the sender claimed in the message and participants' expectations about who would contact~them.

\subsubsection{{URLs and Perceived Convincingness} 
}
Sixteen participants (64\%) discussed the URL when explaining why messages seemed convincing or suspicious. Six described how a URL could provide a sense of trust. For~example, one mentioned how they ``kind of trust'' my university, which was in the URL (T41). Three participants mentioned convincing characteristics with regard to the domain of the URL. One stated, ``The biggest part of this one [URL], and~I even debated whether I would put it first or second; but it has an .edu link . . . and maybe I'm very ignorant about this, but~I feel like that's harder to, like, create a fake .edu website'' (T16). Another participant trusted a URL because it had a .net domain (T41).
In contrast, another participant said they would {not} click on a link because of its untrustworthy domain structure: ``At first, I thought, Oh, I would click on that. But~then it says google.com.org; which is weird, because~I've never seen that before'' (T07). 

Two participants were more likely to click on the link if it included the HTTPS protocol. One said that ``If it looks legit, yeah, like HTTPS and the /[City]Library, something like that'' (T41), then they would click. Another participant said, ``they all say HTTPS, which makes me feel like it's secure'' (T07). Personal association with a URL may also play a role, with~one participant stating they were more likely to click on the link if it included their name within the URL, saying that ``the link actually has my name in it. That would have really like, definitely thrown me for a loop, like really made me like, actually think about clicking the link just because it's like, oh, this is actually from maybe from [my university]''~(T20).

Ten participants were concerned about link shorteners. One mentioned that ``I feel like anytime I've seen a tiny URL address it, like, either hasn't been real, or~it's been like, weird or just different things'' (T16). Six were less likely to click when a link was misspelled, for~example ``facebock'' instead of ``facebook'' (T13). Participants treated the URL name, domain, protocol, spelling, and~use of link shorteners as cues when deciding whether a message appeared~legitimate.

\subsubsection{Technology Communication~Medium}
The communication medium was discussed by 10 participants (40\%). One said it was ``not that abnormal for [them] to receive'' text messages from their home city (T13). Most, however, treated SMS as suspicious when it did not fit the claimed sender. One explained that ``at work we were warned that all, like, valid messages will be through emails'' (T25), while another said, ``This one, [employer] is not going to reach out to me via text for security stuff, period'' (T08). T37 mentioned that they do not have a ``company texting deal.'' Another noted that ``it's very rare that people message me, you know, we talk over Slack or we talk over WhatsApp, maybe through email'' (T31). Others said that an online learning platform (T09) or genealogy website (T02) would not have sent them a text. The~communication channel mattered mainly in relation to the claimed sender: receiving a text could seem appropriate from one source but suspicious from~another.

\subsubsection{Messaging~Style}
Ten respondents (40\%) mentioned issues related to message styling, such as text formatting, tone, and~structure. Four participants acknowledged that style played a role in enhancing the credibility of a message. For~example, T09 affirmed that a message was ``more convincing'' because the formatting and tone felt ``kind of personal.'' T31 also emphasized formatting in relation to message personalization explaining, ``First, they talk about how they found me. They loved my work. And~they saw my music profile. They talk about a music festival.... The~format seems like it's from an organized festival.'' A message's ``casual style'' was also suggested by respondents as a way for the message to feel authentic. A~message's ``less formal'' style was more ``enticing'' and ``attractive'' to T41. In~contrast, two participants mentioned that the message tone was off-putting. For~example, T09 identified a message as potentially a phishing attack due to its ``sales-y'' tone. Additionally, T18 mentioned how sales-oriented ``buzzwords'' indicated a message ``feels more~phishy.''

A quarter of respondents (6 of 25) discussed the role of emojis in the context of message style. All six asserted that emojis diminished the credibility of the message. T34 expressed that ``I think a lot of emojis is kind of something that usually is a red flag for me.'' Multiple respondents conveyed that they did not anticipate receiving emoji-filled messages from senders claiming to represent professional organizations. For~instance, T27 said that in the case of a legitimate organization such as a city biking club, emojis would be ``out of place.'' Similarly, T20 explained that if a message claimed to be from their university, the~presence of emojis would cause them to feel ``thrown off.'' For these participants, emojis in messages claiming to come from professional groups led them to view the messages as less~legitimate.

\subsubsection{Urgency and~Scarcity}
Urgent wording was mentioned by 32\% of participants (8 of 25). Two of the eight mentioned they were more likely to click on a link if it included urgent or fear-inciting language. One explained, ``And because of 'suspicious activity' under my university account, then my first thing is, well, I better click on this. Because~if there's an issue with my account, I want to fix it right away'' (T34). The~other target mentioned the same thing, that they would click on the link because they wanted to fix the problem as soon as possible. They explained, ``And just like, I'd read through it real quick, and~just click on it because I'm like, oh, shoot, that's maybe something important that's going on. Okay'' (T06).

The remaining participants (6 of 25) said they were {less} likely to click if the message used fear or urgency. ``That's always a way of like, okay, they're trying to get you to shut off your logical brain. Put yourself into 'Oh, my gosh, we have to do this right now.' Yeah. Right.'' Another participant confirmed, ``If it seems urgent to you, you're not going to click it because it's very dangerous'' (T13). Similarly, T07 explained, ``I don't like anything where it's like hurry, offer and send because that always makes me wary''. One participant was unsure because a message was ``kind of threatening to lock me out of all of my educational accounts if I didn't do what it asked for'' (T16). Another described a message about a virus affecting young dogs as ``fear mongering'' (T01). Participants reacted differently to urgent language. Some associated it with an important problem requiring action, while others recognized it as pressure intended to discourage careful~evaluation.

\subsubsection{Context~Inaccuracies}
The presence of inaccuracies in the messages was mentioned by 28\% (7 of 25) of the targets. These inaccuracies contributed to perceptions about message credibility. In~this sense, inaccuracies can be understood as failed personalization: the messages used target-specific context, but~in ways that did not match the participant's actual relationships, responsibilities, or~expectations. Participants identified discrepancies related to the names or characteristics of individuals or entities within their professional spheres. For~example, two targets remarked about the message mentioning a colleague that does not exist, stating ``there's no Mike at work'' (T37) or ``there's no one named Sarah on the [my university] instructional design team'' (T18). Another participant expressed confusion about being questioned on matters not relevant to their professional duties, saying that they ``don't deal with payables. So why is that asking me regarding payment?'' (T25). Another target noted discrepancies related to job responsibilities, asserting that such library-related decisions would ``go through me at the library,'' making any logistical deviation ``immediately suspicious'' (T01). Regarding messages related to their hobbies, one participant raised skepticism about the message claiming a ``great deal for a hiking trail'' which the participant knows through experience doesn't actually require payment for use (T30). Another target was also suspicious of the message being from a library that they ``just don't use... so they would not know me to contact me'' (T08). In~summary, the~instances where inaccuracies were mentioned ranged from misrepresented personnel and responsibilities to factual discrepancies about the targets' affiliations, activities, and~expectations.

\subsubsection{Plausible~Rewards}
Seven participants (28\%) described realistic rewards as increasing message credibility. One said they would be more likely to click ``because the reward is connected to something I put so much time and effort into'' (T31), while another described a plausible offer: ``It's like taking you to a link to look at an offer, but~it's not a crazy insane offer'' (T16). T34 called one opportunity both exciting and realistic. Conversely, six participants (24\%) described rewards that seemed ``too good to be true'' as suspicious. T37 said it was ``probably not likely'' to be ``developer of the month four times a year.'' T34 questioned free offers, and~T27 warned, ``Anything that's free is a little too good to be true. Okay, so that's when I would be very careful.'' Participants generally described realistic, context-appropriate rewards as more credible and implausibly generous rewards as warning~signs.

\section{Source Attribution~Results}
This section addresses RQ3, RQ4, and~RQ5 by presenting the human source-attribution results and the computational text-attribution~results.

\subsection{Human Source~Attribution}
This subsection addresses RQ3, which asks how accurately participants identified whether the study messages were GPT-4-generated or~human-authored.

\subsubsection*{Identifying Message~Origin}
Participants correctly identified 156 of the 300 message sources (52.0\%). Because~each participant judged 12 messages, our analysis accounted for repeated judgments by the same person. The~estimated accuracy remained 52.0\%, with~a 95\% confidence interval from 44.6\% to 59.3\%. This interval includes the 50\% expected from guessing between two equally common sources, and~the difference from chance was not statistically reliable ($p = 0.597$). Resampling the 25 targets produced a similar interval of 44.7\% to 59.7\%. We therefore found no evidence that participants identified message source more accurately than~chance. Table~\ref{table_confusion} presents the source-judgment counts.

\begin{table}[H]
\centering
\caption{{Confusion matrix for perceived message source. Participants correctly identified 156 of 300 message sources (52.0\%). After~accounting for repeated judgments, the~95\% confidence interval was 44.6--59.3\%; the text reports the comparison with chance. Shaded diagonal cells show correct source judgments; bold labels identify totals and correct-guess counts.} 
}
\begin{tabularx}{\textwidth}{lCCC}
\toprule
 & \textbf{Guessed GPT-4} & \textbf{Guessed Human} & \textbf{Row Total} \\
\noalign{\hrule height 0.5pt}
\textbf{{True GPT-4}} & \cellcolor{blue!12}\textbf{{78}} & 72 & 150 \\
\textbf{{True Human}} & 72 & \cellcolor{blue!12}\textbf{{78}} & 150 \\
\noalign{\hrule height 0.5pt}
\textbf{{Column Total}} & 150 & 150 & 300 \\
\textbf{{Correct Guesses}} & \multicolumn{3}{c}{\textbf{{156 of 300 (52\%)}}} \\
\bottomrule
\end{tabularx}
\label{table_confusion}
\end{table}

To explore RQ4 quantitatively, we examined whether correct source judgments were associated with emoji presence, modified links, or~message length. The~exploratory model did not find a reliable overall association between these features and correct source judgments ($p = 0.325$). Because~this analysis involved only 25 targets and occurred after participants learned that AI-generated messages were present, it does not show that these features are uninformative in other~settings.

Although the targets described some of their reasoning behind their guesses on which messages were AI-generated, 12 of the 25 participants stated in some way that they were uncertain about their decisions, often relying on intuition rather than any specific criteria. One target remarked, ``sometimes it's a gut feeling maybe more than like a specific thing you're looking for'' (T07), while another said it's ``just a feeling'' (T02). Others attributed their lack of criteria to advancements in AI, saying that they ``don't really, really know my criteria, because~I know that AI is getting so good'' (T04). Several targets openly admitted their lack of expertise, saying they ``have no idea'' (T15), and~``actually I don't know, I don't know'' (T23). Some were unsure if AI could personalize content or include icons in text (T02), while others conveyed concerns about AI competence, hoping that ``the ones that are worse would hopefully be the AI ones'' (T34). Overall, these responses show the difficulty the targets faced in distinguishing between AI- and human-generated content. In~general, they did not seem to have mental models that provided meaningful guidance in identifying AI-generated~messages.

\subsection{Human Criteria for Judging Message~Source}
To address RQ4 qualitatively, we examined the criteria participants described when judging message source. Because~their overall accuracy did not exceed chance, these themes represent perceptions of AI- and human-authored writing rather than validated indicators of source. Table~\ref{table_theme_source} reports the percentage of the 25 participants who mentioned each theme at least once. Participants could mention more than one~theme.

\begin{table}[H]
\centering
\caption{{Themes influencing perceptions of messages as human- or AI-generated.} Although participants used varied criteria to infer message source, the~most common response was~uncertainty.}
\begin{tabularx}{\textwidth}{lR}
\toprule
\textbf{Theme} & \textbf{\% of Participants} \\
\noalign{\hrule height 0.5pt}
{Uncertainty}     & {48\%} \\
Style           & 40\% \\
Personalization & 32\% \\
Word Choice     & 24\% \\
Message Structure & 24\% \\
Grammar         & 24\% \\
Emojis          & 20\% \\
Message Length  & 16\% \\
URL             & 8\% \\
\bottomrule
\end{tabularx}
\label{table_theme_source}
\end{table}
\unskip

\subsubsection{Style}
The message style was mentioned by 40\% (10/25) of the targets. Many targets noted that AI-generated messages tend to be ``pretty formal'' (T21) or ``overly informal or overly formal'' (T09) while others said messages often appear ``extremely generic sounding,'' as if they were created from a prompt with specific parameters (T18). In~contrast, typos and awkward phrasings were often treated as signs of human error rather than AI authorship. Such mistakes contrasted with the ``more robotic'' nature of AI messages described by some participants (T08). Some targets observed that AI-generated messages might seem ``too perfect,'' such as refraining from casual texting language that humans often use, such as ``RN'' for ``right now'' (T34). One participant noted that the excessive use of exclamation points is what made them think the message is AI-generated: ``I mean, honestly, what's driving me insane about all of these is exclamation points. Yeah. I'm like, why are there so many exclamation points all over?'' (T07).

In contrast, messages that gave a sense of urgency were more likely to be perceived as human-authored, as~one target explained: ``Then I didn't put this one because they were actually pressuring me to do it immediately. So I don't think AI can do it'' (T02). Overall, targets indicated that a message's style played a role in their AI-versus-human decisions, with~AI messages being attributed to using a style that is overly formal, generic, or~laden with exclamation points, and~with human messages appearing more casual, urgent, and~imperfect.

\subsubsection{Personalization}
Message personalization was mentioned by 32\% (8/25) of the targets. Five participants mentioned that a lack of personalization made them suspect the message was AI-generated. For~instance, one target noted, ``AI were the ones that were a little bit less personal'' (T34), while another expressed doubt, saying, ``There is no connection with me as a reader'' (T41). Some targets believed the absence of personal pronouns or a personal introduction with the recipient's name indicated AI authorship. One target observed, ``All of these messages, but~two, include my name'' (T37), leading them to suspect those two were AI-generated. Conversely, four participants cited personalized messages as evidence of human authorship, providing similar reasons as those who suspected AI. This group of participants felt messages tailored to the recipient sounded human, with~one target commenting, ``it just sounds very personal'' (T21). Additionally, recipients noted that human-sounding messages often began with an introduction of the sender, such as ``this is Sarah from the [my university] student instructor program'' (T41). Overall, personalized messages were deemed more human-like, while those lacking personalization were associated with~AI.

\subsubsection{Word~Choice}
Message word choice was mentioned by 24\% (6/25) of the targets. Four of the targets claimed that the word choice sounded like AI, particularly the use of buzzwords and marketing words. One target explained that messages sound AI-generated when they use buzzwords that are associated with being ``attention grabbers'' because ``AI would utilize those a lot'' (T16). Another target described a message as ``too specific'' when it included words that humans ``wouldn't necessarily use'' when describing instructional design (T18). A~few of the targets agreed that some of the messages had sentences that a human would have worded differently (T09). Conversely, two targets gave evidence of human-sounding messages and both mentioned slang words. One target described ``There's so many slang terms on it, that it seems really human to me'' (T34), while another defended the messages they thought to be human-authored and explained, ``oftentimes, it was just because I felt like the language seemed a little more slang-like'' (T06). In~short, messages with specific, marketing-like words were perceived as AI-written while messages with casual, slang words that sounded more conversational were perceived as~human-written.

\subsubsection{Message~Structure}
The structure of the message content was mentioned by 24\% (6/25) of the targets. Participants pointed out various aspects of structure that hinted at AI involvement, including repetitive phrases and awkward sentence construction. For~instance, one participant thought a message was AI-generated because ``it looks like there's a template. So there's like a flow that you know, there's a pattern you see it as maybe started by a human'' (T41). Another noted instances where messages seemed generic or mass-produced, akin to ``the email that is sent to everyone in the school'' (T41). One also pointed out inconsistencies within messages, such as repetitive phrases or ``two different things, in~the same message'' (T04). One target also noted that AI-generated messages tended to be more wordy, containing ``lots of filler phrases'' compared to human-authored messages (T37).

\subsubsection{Grammar/Spelling}
The grammar and spelling of the messages were mentioned by 24\% (6/25) of the targets. Four of the targets explained that the grammar and spelling were ``too perfect'' (T34), which led them to believe the message was AI-generated. One described that ``the grammar was almost too correct'' (T06). Messages that were written ``not how you would actually speak'' were considered to be AI-generated (T13). Three of the targets gave evidence of spelling and grammar that seemed human-like. One target said ``this one's a legitimate person right there, because~they spelled the 'util source' wrong'' (T37), while another gave a similar reasoning and stated, ``I don't think that AI would have made the grammar mistake'' (T21). If~the messages included a mix of good grammar with bad spelling or some texting language, a~human author became more believable. While the participants who noted grammar and spelling were split between AI- and human-based evidence, most came to the same conclusion that AI would have perfect grammar and spelling while humans would have made~errors.

\subsubsection{Emojis}
The use of emojis in messages was mentioned by 20\% (5/25) of the targets as being indicative of an AI-generated message. Two concluded that the use of emojis makes a message appear to be written by AI (T25), with~one target explaining, ``there's always an emoji'' and that the emoji messages don't look very personal (T30), indicating a machine wrote the message. Alternatively, three of the targets connected emoji use to human writing because they were not sure if AI could use emojis. One target claimed they didn't think a message was AI-written because of ``the icon'' (T02), while another explained ``I've never asked an AI to do something with emojis. I wasn't sure if it could. I've never seen that, but~I wouldn't expect emojis'' (T18). The~targets' opinions on whether the message was human- or AI-written seemed influenced by their awareness of AI's capability to use emojis and their previous encounters with AI using emojis. In~our spear phishing messages, emojis appeared in 66\% of messages written by AI and only 2\% of messages written by~humans.

\subsubsection{Message~Length}
Message length was mentioned by 16\% (4/25) of the targets. Participants assumed messages made by AI would be longer than human-made messages. One mentioned that they would ``have AI do the longer ones and I would write the shorter ones myself'' (T21), assuming those who created these messages would do the same. These four targets were correct in their hunch that AI messages would be longer. Across the messages, AI-generated messages had on average 41\% higher character count per message, with~337.8 characters for AI and 237.9 characters for human-authored~messages.

\subsubsection{{URLs as Source-Attribution Cues}}
The message URL was mentioned by 8\% (2/25) of the targets. Both of these targets explained that the included URL led them to believe the message was AI-generated. One target mentioned that the word ``dot'' was spelled out rather than typed so it ``seemed like something maybe an AI would do'' (T13). The~other target concluded that altered links, such as Bitly, are evidence of AI and stated that AI ``changed the URL to be something a little bit easier to read and understand'' because they had used {ChatGPT} 
 before and remembered an output with a shortened link. Both targets concluded that a suspicious URL component led to the conclusion that a message was AI-generated. Most (71\%) human authors did not include an actual URL; instead, they included placeholders such as ``[URL],'' ``site,'' or ``url.'' In contrast, only one AI message had a placeholder. As~discussed in the methods, all of these placeholders were replaced with a shortened link from~tinyurl.

\subsection{{Computational Source~Attribution}}
This subsection addresses RQ5 by asking whether the GPT-4-generated and human-authored messages in this study remained distinguishable from their text after progressively stricter~controls.

To determine whether the classifier relied mainly on obvious differences between the two message sets, we repeated target-grouped five-fold cross-validation after jointly standardizing URLs, emojis, case, digits, punctuation, symbols, and~message length. In~the strictest condition, every fold trained on messages from 20 targets and tested on messages from five unseen~targets. Table~\ref{table:joint_normalization} reports the sensitivity results.

\begin{table}[H]
\centering

\caption{{Target-grouped computational source-attribution sensitivity results. Metrics are pooled across out-of-fold predictions. Every fold trained on messages from 20 targets and tested on messages from five unseen targets. The~no-link baseline and normalized conditions were embedded at different times, so differences between them should not be interpreted as the effect of changing any one feature.} 
}
\begin{adjustwidth}{-\extralength}{0cm}
\begin{tabularx}{\fulllength}{lccccc}
\toprule
\textbf{Condition} & \textbf{ROC AUC} & \textbf{Avg. Precision} & \textbf{Bal. Acc.} & \textbf{Sensitivity} & \textbf{Specificity} \\
\midrule
No-link baseline & 0.977 & 0.976 & 0.920 & 0.933 & 0.907 \\
Jointly normalized & 0.961 & 0.960 & 0.910 & 0.913 & 0.907 \\
Jointly normalized and length-equalized & 0.954 & 0.948 & 0.887 & 0.900 & 0.873 \\
\bottomrule
\end{tabularx}%
\end{adjustwidth}
\label{table:joint_normalization}
\end{table}

The GPT-4-generated and human-authored messages remained distinguishable after these controls. In~the strictest condition, balanced accuracy was 88.7\%, ROC AUC was 0.954, and~average precision was 0.948. Across the five groups of unseen targets, ROC AUC ranged from 0.929 to~0.970.

This result shows that the two message sets retained detectable textual differences after the most obvious surface features were jointly controlled. It does not show that these differences are inherent to AI writing or that the classifier would work with other models, prompts, human authors, attacker workflows, or~deliberately revised~text.

\subsubsection*{Descriptive PCA and Exploratory~Clustering}
\begin{figure}[H]
 \hspace{-5pt}   \includegraphics[width=0.88\linewidth,keepaspectratio]{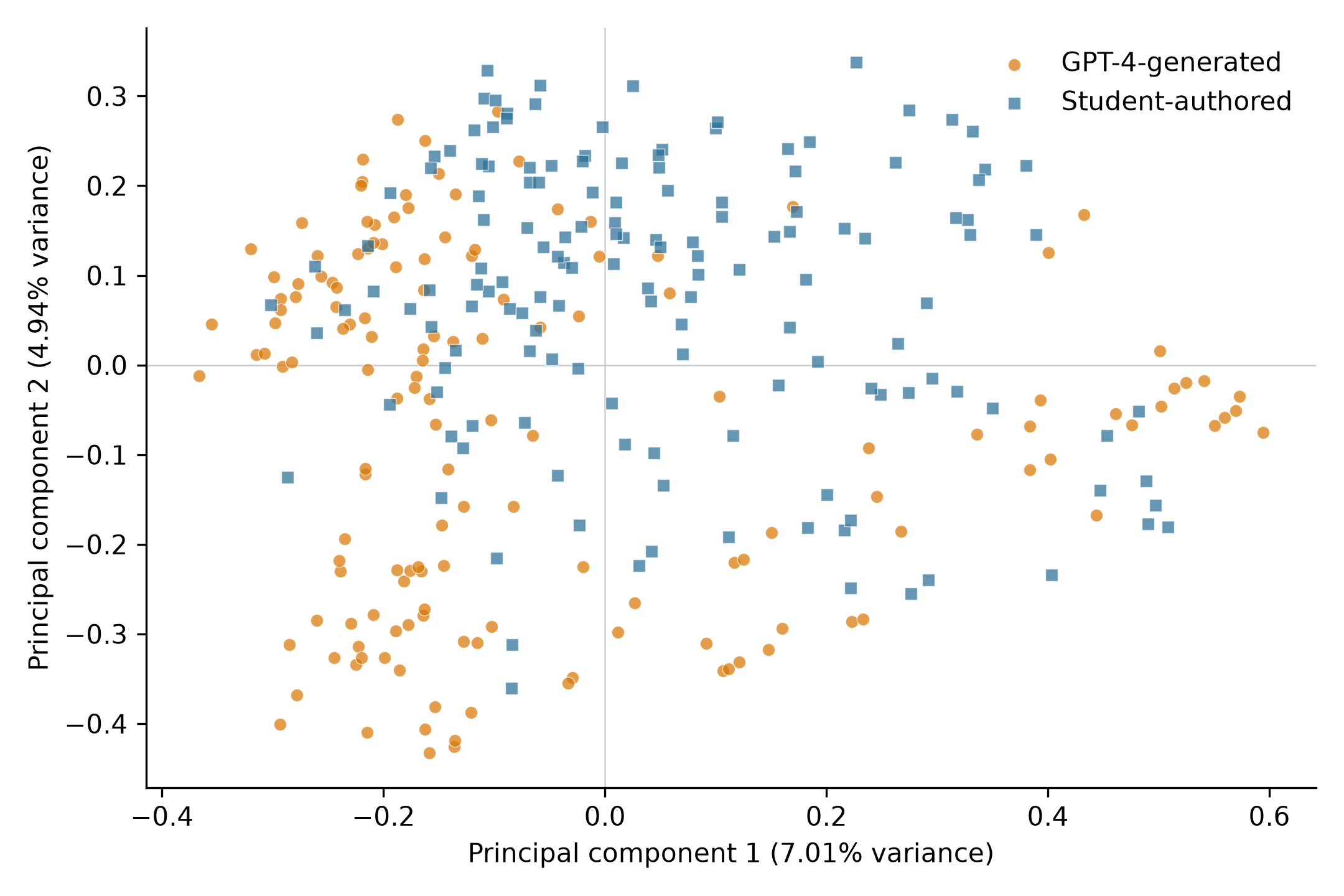}
    \caption{\label{fig:pca_source}{Descriptive PCA projection of the no-link message embeddings by source. The first two components explain 11.95\% of embedding variation. Light gray horizontal and vertical reference lines mark zero on the two principal components. The~projection is included for visualization and is not an independent test of source attribution.} 
}
\end{figure}

To provide an exploratory view of the message representations, Figure~\ref{fig:pca_source} plots the no-link embeddings on their first two principal components. These dimensions explain 11.95\% of the variation in the embeddings, so the figure provides only a limited two-dimensional view and is not used as a statistical~test.

Supplementary Materials, {Section~2}, provides an exploratory KMeans clustering analysis. Because~the selected 13-cluster solution was not uniquely supported and its selection considered source separation, we do not use clustering to support conclusions about message~source.

\section{Discussion}
The discussion follows the study's two analytic tracks before considering the proposed TRAPD framework and broader defensive~implications.

\subsection{Convincingness of GPT-4- and Human-Authored~Messages}
A growing body of work has found that recent LLMs can produce content that equals or exceeds human-authored content in domains such as advertising, moral argumentation, and~political speech~\cite{zhang_human_2023,nisbett_how_2023,palmer_large_2023}. Phishing research has begun to show a similar pattern. Heiding~et~al. found that human V-triad and human+GPT-4 phishing messages produced the highest click-through rates in a field experiment, while GPT-4 still outperformed a control message; more recent work suggests that fully automated LLM-enabled phishing can perform comparably to human-expert phishing in email settings~\cite{heiding_devising_2023,heiding_evaluating_2024}. Hazell likewise argues that LLM-generated spear phishing can be realistic and cost-effective~\cite{hazell_large_2023}.

Our study extends this literature by comparing GPT-4-generated and human-authored spear phishing SMS messages tailored to the same targets. In~this 25-target pilot, participants placed 28.0\% of GPT-4-generated messages and 21.3\% of human-authored messages above their intended-click thresholds. The~estimated difference was 6.7 percentage points, but~the confidence interval ranged from 2.9 points in favor of human-authored messages to 16.3 points in favor of GPT-4-generated messages. The~study therefore does not establish that GPT-4-generated messages were more convincing, nor does it establish that the conditions were~equivalent.

The practical concern is that a simple LLM prompt produced personalized messages that participants found comparable, within~the uncertainty of this pilot, to~messages written by screened novice student authors. If~inexpensive prompts can repeatedly produce messages of comparable quality, the~scale of message generation becomes part of the threat even without evidence that individual AI-generated messages are superior. The~comparison leaves room for improvement on both sides: the LLM prompt was simple, and~the human authors were not trained using the V-triad approach used successfully by Heiding~et~al.~\cite{heiding_devising_2023}.

The topic results add a practical dimension. Participants placed 38\% of job-related messages above their intended-click thresholds, compared with 19\% of hobby messages and 17\% of social-media-related messages. After~accounting for repeated judgments and multiple comparisons, job-related messages had reliably higher intended-click odds than both other topics. The~rankings followed the same descriptive pattern, although~their pairwise differences were not statistically reliable after~adjustment.

This suggests that workplace pretexts deserve particular attention in personalized smishing research and training. The~result is consistent with prior work showing that workplace phishing can exploit role expectations, authority, urgency, and~organizational relevance~\cite{hanus_phish_2022,burns_phishing_2019,williams_exploring_2018}, although~most of that work concerns email rather than SMS. Because~this was a small pilot measuring intended rather than observed clicking, the~size of the topic difference should not be generalized directly to real-world~behavior.

Participants' explanations help show why personalization can increase credibility or raise suspicion. Nineteen of the 25 participants discussed personal relevance, and 17~described accurate connections to their interests or responsibilities as making messages more convincing. At~the same time, 17 participants described unfamiliar or inconsistent senders as warning signs. Participants also flagged incorrect organizations, invented colleague names, irrelevant job responsibilities, and~communication channels that the claimed sender would not normally~use.

These examples show that including personal information does not automatically make a message convincing. The~information must fit the target's actual relationships, responsibilities, and~expectations. When it does not, personalization can fail and expose the message as suspicious~\cite{oliveira_empirical_2019,karamagi_review_2021,butavicius_breaching_2016,stembert_study_2015}.

\subsection{Source Attribution by Participants and~Models}
The source-attribution results reveal an important contrast. Participants correctly identified message source 52.0\% of the time, which did not differ from chance. They described cues such as formality, word choice, structure, grammar, emojis, and~message length, but~these cues did not lead to reliable source judgments. Their explanations are useful for understanding beliefs about AI writing, not as validated indicators of authorship. This aligns with prior work showing that people often struggle to distinguish AI-generated from human-authored language~\cite{kobis_artificial_2020,jakesch_human_2023}.

The computational analysis detected patterns that participants did not use reliably. The~GPT-4-generated and human-authored messages remained distinguishable when the classifier was tested on messages from unseen targets, even after URLs, emojis, case, digits, punctuation, symbols, and~message length were jointly standardized. This suggests that the two study-specific message sets differed in patterns distributed across the text rather than only in their most obvious surface~features.

The result remains limited to one GPT-4 condition, one prompt design, and~one pool of novice student authors. It is not a general AI detector and may not extend to other settings. Writers can also deliberately revise AI-generated text to avoid detection; Cheng~et~al. found that detector-guided paraphrasing reduced the effectiveness of several AI-text detectors~\cite{cheng_adversarial_2025}. The~classifier should therefore be treated as evidence about this message set, not as a reliable basis for blocking messages in~practice.

\subsection{{The Proposed TRAPD Evaluation~Framework}}
TRAPD is best understood as a middle-ground approach between survey-style message evaluation, laboratory phishing-judgment tasks, and~deceptive field experiments. Compared with one-shot survey judgments, TRAPD gives participants multiple personalized messages, asks them to rank those messages, mark an intended-click threshold, and~explain their reasoning. Compared with laboratory detection tasks such as PEST~\cite{hakim_phishing_2021}, it emphasizes within-target comparison of personalized deceptive content. Compared with field experiments, it sacrifices ecological realism while preserving informed consent and enabling grounded explanations that are difficult to obtain after a single live~exposure.

Field experiments are stronger for measuring real-world click behavior, while TRAPD is stronger for controlled comparison, ethical administration, and~explanation of target reasoning. Prior work suggests that survey and laboratory methods can capture meaningful phishing variables such as convincingness, suspicion, and~intended response~\cite{thomopoulos_methodologies_2023,hakim_phishing_2021}, while contextual factors can influence real-world behavior~\cite{distler_context_2023}.

TRAPD should be viewed as a proposed evaluation framework rather than a validated measurement instrument. In~this study, it supported internally valid comparisons between messages written for the same target under controlled conditions. However, the~study did not test whether TRAPD produces consistent results over time, whether different evaluators apply it consistently, whether changing message order affects the results, or~whether its intended-click threshold corresponds with clicking in a live setting. Comparing TRAPD directly with survey, laboratory, and~authorized field methods is an important next~step.

Several implementation lessons may help future studies. Sorting 12 items took approximately 10--15 min, and~we would not recommend exceeding that number. The~delayed follow-up contributed to attrition, so future studies should make the later session clear during recruitment and schedule it when possible at the initial survey. URL construction should also be standardized across~conditions.

\textls[-15]{Increasing the sample after data collection was impractical because the human-authored messages had already been created for specific targets. A~fully automated study could instead generate personalized content, collect rankings and explanations, and~compare LLMs or prompt designs within one online workflow. This could support larger samples while preserving TRAPD's focus on controlled comparison and participant~reasoning.}

Any automated version should be limited to authorized research, evaluation, or~training and should minimize the personal information collected, preserve informed consent, protect target information, and~prevent operational misuse. Research on large-model agents suggests that future studies may also need to examine more complex, multi-step AI-supported workflows, although~that question goes beyond the single-message generation tested here~\cite{wang_agents_2026}.

\subsection{Implications of AI-Enabled Spear~Phishing}
Generative AI is widely adopted in organizations, with~a 2024 McKinsey \& Company survey reporting that 72\% of organizations had adopted AI in at least one business function~\cite{mckinsey_state_of_ai_2024}. That adoption brings legitimate benefits, but~malicious actors may also use fluent, low-cost LLM output to create personalized phishing content. Existing cybercrime law and accountability frameworks may not fully address these risks~\cite{raja_ai_accountability_2023}.

The risk examined here comes from the combination of comparable message quality within this pilot's uncertainty, low-cost generation, and~participants' inability to identify source more accurately than chance. First, a~simple GPT-4 prompt generated personalized messages that participants found comparable to messages written by screened novice student authors. Second, the~study-specific texts contained computationally detectable source patterns even though participants could not use their perceived cues reliably. Prior work has examined phishing detection and machine-learning frameworks~\cite{basit_survey_2021,zieni_phishing_2023,dou_systematic_2017}, but~less is known about how defenses should adapt when personalization and fluent language become~inexpensive.

\subsection{Dual-Use~Considerations}
The procedures used to study personalized deception could also be misused to improve deceptive messages or automate targeting. This study is intended to help researchers understand risk, evaluate defenses, and~improve training, but~those goals do not remove the need for safeguards. Tools that automate message generation and targeting could make misuse easier, while personalized messages, participant records, and~message embeddings may reveal sensitive information. Automated TRAPD studies should therefore be limited to authorized settings with informed consent, collection of only the information needed, access controls, and~review of generated content. Our data-sharing plan releases aggregate findings and analysis code while restricting personalized data, embeddings, and~operational generation~materials.

\subsection{Mitigation and Countermeasure~Implications}
Participants did not identify message source more accurately than chance, and~the cues they described were inconsistent. Training should therefore emphasize contextual verification rather than whether a message ``sounds like AI'': whether the sender, communication channel, request, and~URL match the recipient's~expectations.

Because job-related messages were more convincing than hobby- or social-media-related messages in this study, awareness efforts may benefit from realistic workplace scenarios, especially those involving urgent tasks, account problems, or~requests that appear relevant to the target's role. Personalized training should also make clear that personalization itself is not evidence of~legitimacy.

Although this was an unsolicited qualitative observation rather than a measured training outcome, several participants reported that the study made them more aware of how convincing personalized AI-generated phishing could be. Future research could test whether realistic personalized examples improve verification behavior, particularly checking the sender, communication channel, URL, and~request rather than guessing whether a message sounds human- or~AI-authored.

The computational results suggest a possible role for automated analysis, but~not a ready-made detector. Any practical tool would require validation across broader settings in which attackers can adapt. The~more immediate defensive lesson is that controls should assume attackers may combine plausible context, fluent writing, and~personalization regardless of~source.

\section{Limitations}
This 25-target pilot provides a controlled comparison of personalized messages, but~the following limitations define what its findings can and cannot~show.

\subsection{Measurement and TRAPD~Validity}
TRAPD measured perceived convincingness and intended clicking rather than responses to live SMS messages. Participants completed the task in an interview using printed messages, without~delivery details such as sender number, timing, device context, distractions, or~functional links. These choices supported controlled comparison, informed consent, and~detailed explanations, but~reduced realism. Perception and intended-response measures are commonly used in phishing research~\cite{hoeken_perceived_2022,moody_phish_2017,zhuo_human-centered_2023,hakim_phishing_2021}, but~they are not equivalent to observed~behavior.

TRAPD is also a proposed evaluation framework rather than a validated measurement instrument. This study did not test whether it produces consistent results over time, whether different evaluators apply it consistently, whether message order changes results, or~whether its intended-click threshold predicts live clicking. The~findings should therefore be interpreted as controlled comparisons of participant judgments rather than estimates of real-world click-through~rates.

\subsection{Sample and Statistical~Precision}
The study included 25 targets, which limits the precision and generalizability of the findings. Recruitment relied partly on convenience sampling, and~25 of the 41 people who completed the initial survey (61\%) returned for the interview. Although~the final group varied in age, gender, university affiliation, and~profession, it should not be treated as representative of a broader~population.

The confidence interval around the source difference was wide enough to include effects in either direction. The~study therefore establishes neither superiority nor equivalence and was too small to support conclusions that demographic or prior-experience differences were absent. For~future planning, simulations under the present design assumptions suggest that approximately 100 completed targets would provide 80\% power to detect a 6.7-percentage-point difference. Detecting a more conservative 5-percentage-point difference would require approximately 165 completed targets. These estimates guide future recruitment and are not post-hoc power estimates for this completed~study.

\subsection{Human Comparison~Condition}
The human-authored messages were written by time-constrained novice student authors who had received phishing-related instruction. Their messages were screened for minimum quality, reviewed by a team that included cybersecurity professors, and~randomly selected from the valid pool. These procedures strengthened the comparison condition, but~the authors were not professional social engineers, security testers, or~a representative sample of human attackers. Giving both conditions the same prompt improved consistency without making the human and GPT-4 generation processes equivalent. The~findings apply to this specific comparison and should not be generalized to AI versus humans~overall.

\subsection{GPT-4 Generation~Reproducibility}
The evaluated messages and available generation outputs have been preserved, allowing the final analyses to be reproduced. The~batch-generation notebook records {the exact API model alias, \texttt{gpt-4},} 
 temperature 0.7, maximum output length of 256 tokens, and~three generations per prompt. The~available outputs account for all 150 GPT-4-generated messages evaluated in the~study.

However, the~exact model snapshot, API logs, complete software environment, a~single script recreating the entire generation process, and~unspecified parameter defaults were not recorded. The~original generation run therefore cannot be recreated exactly. Future studies should preserve the exact model and access date, all settings and prompts, retry and filtering events, output-selection decisions, and~software~versions.

\subsection{Qualitative~Coding}
Each transcript was assigned to one primary coder, after~which the research team reviewed themes and resolved differences through discussion. The~transcripts were not independently coded in full by a second coder, and~no formal inter-coder agreement statistic was calculated. We therefore cannot quantify how consistently another coder would have assigned the themes. The~qualitative findings should be understood as a consensus-based descriptive analysis of participants'~explanations.

\subsection{Interview Sequence and Response~Effects}
Participants ranked the messages and set intended-click thresholds before learning that AI-generated messages were included. The~disclosure therefore could not affect those earlier judgments, but~it may have influenced later source labels and explanations. Because~the message set contained equal numbers of GPT-4- and human-authored messages, participants may also have assumed that the sources were equally common. The~interviewer and participants' awareness that they were being studied may have encouraged answers they thought the researchers expected. Future studies could test different task orders and source proportions, separate disclosure conditions, and~collect later responses privately on a~computer.

\subsection{Computational Scope and Data~Availability}
The computational source-attribution findings are limited to one balanced message set produced with one GPT-4 condition, one prompt design, and~one pool of novice student authors. Testing on unseen targets reduced the possibility that the classifier learned target-specific information, while jointly standardizing obvious surface features reduced the possibility that it relied only on formatting differences. These safeguards strengthen the study-specific result but do not show that the classifier will work with other models, prompts, authors, domains, or~deliberately revised~messages.

The PCA figure preserves only 11.95\% of variation in the embeddings and is included as a descriptive illustration. The~supplementary 13-cluster analysis is likewise exploratory and does not independently validate source attribution. Personalized messages, participant-level data, historical notebooks, transcripts, recordings, and~embeddings will not be released because of privacy, re-identification, ethical, and~dual-use concerns. The~accompanying reproducibility package provides final analysis code and aggregate outputs, but~restricting the raw materials limits independent reproduction from the original~inputs.

\section{Conclusions}
Personalized deceptive messages can be highly convincing, and~current LLMs make them easier to generate at scale. This 25-target pilot used the proposed TRAPD evaluation framework to compare personalized spear phishing SMS messages generated by GPT-4 with human-authored messages written by time-constrained novice student authors. Although~the observed intended-click rate was higher for GPT-4-generated messages, the~difference was not statistically reliable. The~study therefore establishes neither that GPT-4 outperformed the human-authored condition nor that the conditions were equivalent. The~practical finding is that a simple prompt produced personalized messages that participants found comparable, within~the uncertainty of this pilot, to~messages from a screened pool of novice student~authors.

Job-related messages were especially important. Participants placed 38\% of job-related messages above their intended-click thresholds, compared with 19\% of hobby messages and 17\% of social-media-related messages, and~intended-click odds for job messages were significantly higher after adjustment. The~qualitative findings help explain this pattern: participants evaluated messages through relevance, sender identity, URLs, communication medium, style, plausible rewards, and~contextual accuracy. Incorrect organizations, invented colleagues, unfamiliar senders, and~inappropriate communication channels were examples of failed personalization. Workplace pretexts therefore deserve particular attention in future research and~training.

Participants did not identify message source more accurately than chance. In~contrast, the~GPT-4-generated and human-authored messages in this study remained computationally distinguishable when the classifier was tested on messages from unseen targets and obvious surface differences were standardized. This contrast suggests that the study messages contained patterns participants did not use reliably. However, the~result applies only to this message set and should not be treated as a general AI detector. It may not extend to other models, prompts, human authors, or~deliberately revised~text.

TRAPD supported internally valid comparisons between multiple messages written for the same target while preserving informed consent and collecting detailed explanations. However, it does not measure live clicking and has not been formally tested for reliability or validity against real-world behavior. Future studies should use larger and more diverse samples, fully document model-generation settings, compare TRAPD with authorized laboratory or field methods, and~evaluate defensive interventions while protecting personalized data and limiting dual-use risk. Defenses should assume that convincing personalized phishing can be produced with or without AI and teach people to verify the sender, communication channel, URL, and~request rather than guess who or what wrote the~message.

\vspace{6pt} 

\supplementary{The following supporting information can be downloaded at: \linksupplementary{s1}. {Supplementary Materials PDF: Section 1, Earlier No-Link Computational Analysis, including Table A1, Earlier no-link computer-analysis results; Figure A1, ROC curve for the earlier no-link classifier; Table A2, Earlier simple-feature comparisons; and Figure A2, Predicted source probabilities for the earlier no-link classifier; Section 2, Exploratory KMeans Clustering, including Figure A3, Exploratory KMeans clustering shown in two dimensions, and Figure A4, Distribution of GPT-4-generated and human-authored messages across the exploratory 13-cluster solution; and Section 3, Prospective Sample-Size Planning. Also included are a reproducibility package containing deidentified behavioral outcomes, final analysis code, aggregate outputs, workflow documentation, and a prompt template containing only fictional or placeholder values.} 
}




\authorcontributions{Conceptualization, J.F. and D.H.; methodology, J.F., D.H. and B.S.; software, J.F., M.T. and R.C.; validation, J.F.; formal analysis, J.F., D.H., B.S., M.T., S.V.M., R.C. and G.S.; investigation, J.F., D.H. and M.T.; resources, J.F., D.H. and M.T.; data curation, J.F., D.H. and M.T.; writing---original draft preparation, J.F., D.H., B.S., M.T., S.V.M. and R.C.; writing---review and editing, J.F., D.H., B.S., S.V.M. and R.C.; visualization, J.F.; supervision, D.H. and B.S.; project administration, J.F., D.H. and B.S. All authors have read and agreed to the published version of the manuscript.}

\funding{This research received no external~funding.}

\institutionalreview{The study was approved by the Institutional Review Board of Brigham Young University (IRB2023-065).}

\informedconsent{Informed consent was obtained from all participants involved in the study.}

\dataavailability{The raw participant data, personalized messages, interview materials, historical notebooks, and~derived message embeddings are not publicly available and are not available upon request because no external controlled-access process has been confirmed and the materials contain or encode participant-specific information. Deidentified behavioral outcomes, final analysis code, aggregate results, supplementary computational outputs, a~prompt template containing only fictional or placeholder values, and~workflow documentation are supplied as supplementary material with this submission and will be archived in Zenodo upon article acceptance. Questions about data availability may be directed to the corresponding authors.}

\acknowledgments{The authors thank the study participants and student message authors for their time and contributions. During~this study, the~authors used {OpenAI GPT-4 via the OpenAI API (recorded model alias: \texttt{gpt-4})} 
 to generate experimental spear phishing messages. During manuscript revision, the authors used OpenAI ChatGPT and OpenAI Codex for language editing, clarity, organization of reviewer responses, drafting candidate revisions, code review, citation checking, formatting, and figure preparation. The authors reviewed and verified all AI-assisted material and take full responsibility for the content of this~publication.}

\conflictsofinterest{The authors declare no conflicts of~interest.} 

\durcstatement{{This} 
 research concerns AI-enabled social engineering and therefore has dual-use potential. The~study was conducted with informed consent, used simulated messages rather than live phishing deployment, and~reports findings to support defensive research and training. The~authors do not release personalized message data, participant information, derived embeddings, or~operational automation that could increase misuse or re-identification risk. Proposed automated TRAPD studies and tools should be limited to authorized research, evaluation, and~training environments with appropriate ethical and privacy controls.}

%
%



\abbreviations{Abbreviations}{
The following abbreviations are used in this manuscript:\par\medskip

\noindent
\begin{tabular}{@{}ll}
AI & Artificial intelligence\\
LLM & Large language model\\
SMS & Short message service\\
TRAPD & Threshold Ranking Approach for Personalized Deception
\end{tabular}
}

\begin{adjustwidth}{-\extralength}{0cm}

\reftitle{References}

\PublishersNote{}
\end{adjustwidth}
\end{document}